\documentclass{jfm}
\usepackage{etex}
\usepackage{mathtools}
\usepackage{lmodern}
\usepackage{gensymb}
\usepackage[dvips]{graphicx}
\usepackage{multirow}
\usepackage{url}
\usepackage{setspace}
\usepackage{natbib}
\usepackage[percent]{overpic}
\usepackage{graphics}
\usepackage{graphicx}
\usepackage{amssymb}
\usepackage{amsmath}
\usepackage{mathrsfs}
\usepackage{mathtools}
\usepackage{cancel}
\usepackage{soul}
\usepackage{subfig}
\usepackage{float}
\usepackage{textcomp}
\usepackage{listings}
\usepackage{psfrag}
\usepackage{placeins}
\usepackage{color}
\usepackage{epsfig}
\usepackage{bm}
\usepackage{lscape}
\usepackage[usenames,dvipsnames,table]{xcolor}
\usepackage{tikz}
\usetikzlibrary{positioning,arrows}
\usetikzlibrary{decorations.pathreplacing}
\usepackage{pstricks}
\usepackage[toc,page]{appendix}
\usepackage{upgreek}
\usepackage{caption}
\usepackage{multirow}
\usepackage[mathlines]{lineno}
\usepackage{cleveref}
\allowdisplaybreaks
\usepackage[normalem]{ulem}

\usepackage{paralist}

\ifCUPmtlplainloaded \else
  \checkfont{eurm10}
  \iffontfound
    \IfFileExists{upmath.sty}
      {\typeout{^^JFound AMS Euler Roman fonts on the system,
                   using the 'upmath' package.^^J}%
       \usepackage{upmath}}
      {\typeout{^^JFound AMS Euler Roman fonts on the system, but you
                   dont seem to have the}%
       \typeout{'upmath' package installed. JFM.cls can take advantage
                 of these fonts,^^Jif you use 'upmath' package.^^J}%
      }
  \else
  \fi
\fi

\ifCUPmtlplainloaded \else
  \checkfont{msam10}
  \iffontfound
    \IfFileExists{amssymb.sty}
      {\typeout{^^JFound AMS Symbol fonts on the system, using the
                'amssymb' package.^^J}%
       \usepackage{amssymb}%
       \let\le=\leqslant  
       \let\ge=\geqslant  
      }{}
  \fi
\fi

\ifCUPmtlplainloaded \else
  \IfFileExists{amsbsy.sty}
    {\typeout{^^JFound the 'amsbsy' package on the system, using it.^^J}%
     \usepackage{amsbsy}}
    {}
\fi

\newcommand\p{\ensuremath{\partial}}

\newcommand{\beq}{\begin{equation}}
\newcommand{\eeq}{\end{equation}}
\newcommand{\uvec}{\vec{u}} 
\newcommand{\bnabla}{\mbox{\boldmath $\nabla$}}

\newcommand{\Rone}[1]{{#1}} 
\newcommand{\Rtwo}[1]{{#1}} 
\newcommand{\Rthree}[1]{{#1}} 

\newcommand{\arXiv}[1]{}

\renewcommand{\vec}[1]{\mbox{\boldmath $#1$}}

\newcommand{\Nu}{\mbox{\it Nu}}

\newcommand{\pd}[1]{\partial_{#1}}
\newcommand{\T}{\Theta}

\title[A vertical heated pipe]{Suppression of turbulence and 
travelling waves 
in a vertical heated pipe}

\author[E. Marensi, S. He and A. P. Willis]%
{Elena Marensi$^{1,2,}$\thanks{Email address for correspondence: elena.marensi@ist.ac.at}, Shuisheng He$^3$ and Ashley P. Willis$^1$}

\affiliation{$^1$School of Mathematics and Statistics, University of Sheffield, Sheffield S3 7RH, UK\\
             $^2$IST Austria, Am Campus 1, 3400 Klosterneuburg, Austria \\
             $^3$Department of Mechanical Engineering, University of Sheffield, Sheffield S1 3JD, UK
}

\pubyear{}
\date{?; revised ?; accepted ?. - To be entered by editorial office}
\begin{document}

\maketitle

\begin{abstract}

Turbulence in the flow of fluid through a pipe can be suppressed by buoyancy forces. As the suppression of turbulence leads to severe heat transfer deterioration, this is an important and undesirable phenomenon in both heating and cooling applications. \Rtwo{Vertical flow is often considered, as the
axial buoyancy force can help drive the flow.  
With heating measured by the buoyancy parameter $C$, 
our direct numerical simulations show that shear-driven turbulence may either be completely laminarised or transitions to a relatively quiescent convection-driven state.
Buoyancy forces cause a flattening of the  base flow profile, 
which in isothermal pipe flow has recently been linked to complete suppression of turbulence (K\"{u}hnen {\emph{et al.}}, {\emph{Nat. Phys.}}, vol. 14, 2018, pp. 386--390), and the flattened laminar base profile has enhanced nonlinear 
stability  (Marensi {\emph{et al.}}, {\emph{J. Fluid Mech.}}, vol. 863, 2019, pp. 50--875).
In agreement with these
 findings, the 
nonlinear lower-branch travelling-wave solution analysed here, which is believed to mediate transition to turbulence in isothermal pipe flow, is shown to be suppressed by buoyancy. 
A linear instability of the laminar base flow is responsible for the appearance  
of the relatively quiescent convection driven state for $C\gtrsim 4$ across
the range of Reynolds numbers considered.
In the suppression of turbulence, however, 
i.e.\ in the transition {\emph {from}} 
turbulence,
we find clearer association with the analysis of 
He {\emph{et al.}} ({\emph{J. Fluid Mech.}}, vol. 809, 2016, pp. 31--71)
than with the above dynamical systems approach, which describes better
the transition {\emph{to}} turbulence.
The laminarisation criterion He {\emph {et al.}}\ propose,}
based on an apparent Reynolds number of the flow as measured by its driving pressure gradient, is found to capture the critical $C=C_{cr}(Re)$ above which the flow will be laminarised or switch to the convection-driven type. 
Our analysis suggests that it is the weakened rolls, rather than the streaks, which appear to be critical for laminarisation.
\end{abstract}

\section{Introduction}
\label{sec:intro}
Most energy systems rely on fluids to transfer heat from one device to another to facilitate power generation, provision of heating or production of chemicals. Flows are often forced through channels or arrays of pipes taking heat away from the surfaces. In a nuclear reactor, for example, the reactions occur within the fuel pins, which are cooled by flow of coolant through the channels formed by arrays of fuel pins to maintain their temperature within a specific limit as well as transferring energy to the steam generators. 
In an isothermal flow, the volume flux is driven by an externally applied pressure gradient, and the flow is referred to as `forced'. In a vertical configuration, however, buoyancy resulting from the lightening of the fluid close to the heated wall can provide a force that partially or fully drives the flow, referred to as mixed or natural convection, respectively. When heat flux is very high, we can have a `supernatural' state of flow, where the buoyancy is sufficiently strong that 
\Rtwo{a reversed pressure gradient may be necessary to limit 
or maintain a constant volume flux.} 
Under certain conditions (e.g. the Boussinesq approximation) an upward heated flow may be considered equivalent to a downward flow cooled at the boundary (Appendix A).

Mixed convection is of significant importance to engineering design and safety considerations and as such extensive research has been carried out to develop engineering correlations \citep{jackson-etal-1989, yoo-2013}, turbulence models \citep{kim-etal-2008, bae-2016} and a better understanding of the physical flows \citep{you-etal-2003}. 
A particularly interesting physics is that the flow, at a Reynolds number 
where shear-driven turbulence is ordinarily observed, in the presence of buoyancy
 may be partially or fully laminarised, or becomes a convection-driven turbulent flow (i.e. natural convection, referred to above).  Heat transfer may be significantly impaired under such conditions.   
\cite{he-etal-2016} (hereinafter referred to as HHS) modelled the effect of buoyancy 
using a prescribed body force, with linear or step radial dependence, without solving the energy equation. They attributed the suppression of turbulence to a reduction in the apparent Reynolds number of the flow, as measured by the pressure gradient required to drive the flow.  Thus, the forced flow
is compared with the unforced 
``equivalent pressure gradient'' reference flow.

Meanwhile, in ordinary (isothermal) pipe flow, \citet{kuhnen-etal-2018a}, 
observed relaminarisation
attributed to flattening of the base flow profile.
\Rtwo{The idea of flattening was first suggested by \citet{hof-etal-2010} who
showed that when two puffs were triggered too close to each other the downstream puff would collapse due to the flattened streamwise velocity profile induced by the upstream puff.}
In the experiments of \citet{kuhnen-etal-2018a} the flattening was introduced by a range of internal and boundary flow
manipulations and a full collapse of turbulence  was obtained for Reynolds numbers up to $40\,000$.
\citet{marensi-etal-2019} showed the complement effect, i.e. the enhanced nonlinear stability of the laminar flow.
 They found that the minimal seed (smallest amplitude disturbance)
for transition is `pushed out' from the laminar state to larger amplitudes when
the base flow is flattened, thus implying
that a flattened base profile is more stable than the parabolic profile.
Here, buoyancy forces also have a flattening effect 
and turbulence may be partially or fully suppressed.
Furthermore, 
early experimental observations \citep{hanratty-etal-1958,kemeny-somers-1962,scheele-hanratty-1962} and subsequent linear 
\citep{yao-1987a, yao-1987b, yao-rogers-1989,chen-chung-1996, su-chung-2000}
and weakly-nonlinear \citep{rogers-yao-1993,khandelwal-bera-2015} stability analyses suggested that,
for sufficiently large heating, the flow
becomes unstable and transitions to 
a new non-isothermal equilibrium state
characterised by large-scale regular motions.
In agreement with the experiments, the linear theory showed that this instability can occur at low Reynolds number (below 100) and for $Re>50$ the critical value of the Rayleigh number is almost independent of $Re$ \citep{yao-1987a}. The first azimuthal mode was found to be the least stable \citep{yao-1987a, su-chung-2000}, consistent with the double-spiral patterns observed experimentally \citep{hanratty-etal-1958} and the instability was linked to the inflectional velocity profile in the buoyancy-assisted case. As suggested by \citet{su-chung-2000}, a competition between different mechanisms -- driven by either shear or convection -- thus exists and understanding its effect on the nature of the flow is the object of our study.

In particular, in this work, we are interested in whether a flow is turbulent or laminar under certain heating conditions and when a turbulent flow may be laminarised or vice versa under the influence of buoyancy.
\Rtwo{
We address this question for a vertically heated pipe,
initially in the dynamical systems context through linear stability and by investigating how travelling wave solutions are affected by the buoyancy force.  
Next, the nature of the laminarisation
is considered.  In isothermal flow at transitional Reynolds numbers, 
the shear-driven state is known to be metastable -- the probability of 
laminarisation follows a Poisson process with a
laminarisation rate that depends on the Reynolds number.  
In any practical setting, where a pipe is of finite length, its 
length affects the probability of 
turbulence surviving to the end of a pipe.  Hence a range of Reynolds 
numbers for transition are quoted, typically 
between 2000 and 2300.
Therefore, we do not attempt to quantify the full statistical nature of the 
transition in the heated case, but instead we 
focus on the phenomenological-based `equivalent pressure-gradient'
analysis of HHS.}  
Through the above approaches, i.e.\ linear stability, nonlinear travelling-wave and `equivalent pressure-gradient' analyses, we aim to elucidate the physical mechanisms underlying the buoyancy-suppression of turbulence, illustrating the bistability nature of such flows.

\subsection{Nonlinear dynamical systems view}

In subcritical wall-bounded shear flows, turbulence arises despite the linear stability of the laminar state \citep{DR04, schmid-henningson-2001}.
\Rtwo{The implication is that the observed transition scenario can only be triggered by finite amplitude disturbances.}
In the last 30 years our understanding of transition to turbulence in such flows has greatly benefited from a
fully nonlinear geometrical approach which adopts concepts from the dynamical systems theory. 
In this view, the
flow is analysed as a huge (formally infinite-dimensional) dynamical system in which the flow state evolves along
a trajectory in a phase space populated by various 
 invariant solutions, travelling waves (TWs) and periodic orbits (POs).
%
\Rtwo{ Nonlinear travelling wave solutions 
were first discovered numerically
 in the early 1990s for plane Couette flows \citep{n90} and in the 2000s for pipe flows \citep{FE03, WK04, Pringle07}.
Since then, partly thanks to the advances in our computational and experimental capabilities, a growing amount of evidence has been collected for their dynamical importance \citep[see reviews][]{Kerswell05, ESHW07, KaUhlVeen12,graham2021exact}.}
These solutions, often referred to as ``exact coherent states/structures'' (ECSs), 
are believed to act as {\emph{organising centres}} \citep{W01} in phase space, in the sense that, when the flow state approaches them, spatio-temporally organised patterns (streaks and streamwise rolls) 
 are observed \citep{science04, KeTu06}.

ECSs are finite-amplitude non-trivial solutions disconnected from the laminar state and 
enter via saddle-node bifurcations as the flow rate is increased.  Some solutions, typically those 
of higher spatial symmetry, exist at flow rates much below that at which transition is usually observed \citep{Pringle09}.  ECSs are linearly unstable, although with only a few unstable directions.  They may be divided into `upper-branch' and `lower-branch' states, depending on whether they are associated with a high or low friction factor.
Lower branch solutions
are representative of the laminar-turbulent boundary 
\Rtwo{ -- the so called ``edge of chaos" \citep{IT01, SchEck06} --
which separates initial conditions that lead to turbulence from those that decay and relaminarise.}
\Rtwo{The edge comes closest to the laminar equilibrium  at the ``minimal seed'' for transition \citep{kerswell-2018}.}
Lower-branch solutions are believed to mediate the transition to turbulence \citep{duguet07, SchEckYor07}, while some upper-branch solutions are embedded in the turbulent attractor and are representative of the turbulent dynamics \citep{AvMeRoHo13, WFSBC15}.

Here, we are interested in studying how travelling wave solutions are affected by the buoyancy force in a vertical heated pipe, and, in analysing their dynamics, we aim to elucidate the physical mechanisms underlying the buoyancy-suppression of turbulence.
%
The transition between regimes is first investigated using linear stability in \S \ref{sec:LS},
followed by analysis of travelling waves in \S \ref{sec:N4L}.

\subsection{Equivalent pressure-gradient (EPG) analysis of HHS}
\label{sect:HHSoverview}

\Rtwo{
Rather than simulating a temperature field, to reduce complexity
HHS considered a fixed radially-dependent axial body force that models the 
buoyancy force, and applied this to isothermal flow.
Conventionally, heated flows are compared with the isothermal (unforced) 
flow at equivalent flow rate (EFR), but HHS observed better comparison with flows at the equivalent pressure gradient (EPG).}
\Rtwo{  In particular, 
after careful analysis,}
%
they observed that adding the radially-dependent force does not alter the turbulent viscosity of an unforced flow 
driven by the same pressure gradient
\Rtwo{(see figure 10 therein)}.  
\Rtwo{The unforced EPG flow is therefore a reference 
flow for cases with the extra radially-dependent forcing.}

\Rtwo{
Note that in a fixed mass-flux calculation, the pressure gradient reduces in 
response to driving from the buoyancy.
Given a heated flow at a particular Reynolds number $Re$
(defined in terms of the mass flux), 
to determine the Reynolds number of the EPG flow, one must split 
the mass flux into contributions from the pressure gradient
and from the buoyancy.  
The former component determines the
`apparent Reynolds number' $Re_{app}$ of the EPG flow.}
Laminarisation of the body forced flow is observed to 
\Rtwo{occur when its $Re_{app}$ is consistent with the $Re$ 
at which laminarisation occurs in isothermal flow.}
Further details of the analysis
are provided in \S \ref{sec:HHS} and HHS prediction is compared with a suite of 
simulations in \S \ref{sec:HHSpred}.

\section{Formulation}
\label{sec:formulation}

\begin{figure}
\centering
\includegraphics[width=0.3\textwidth]{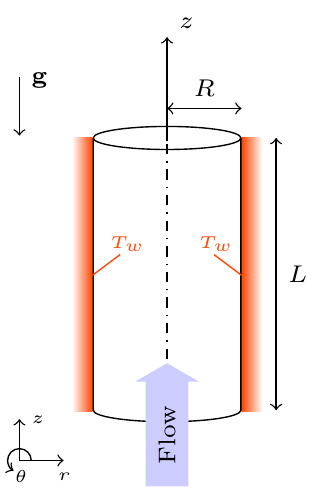}
\includegraphics[width=0.6 \textwidth]{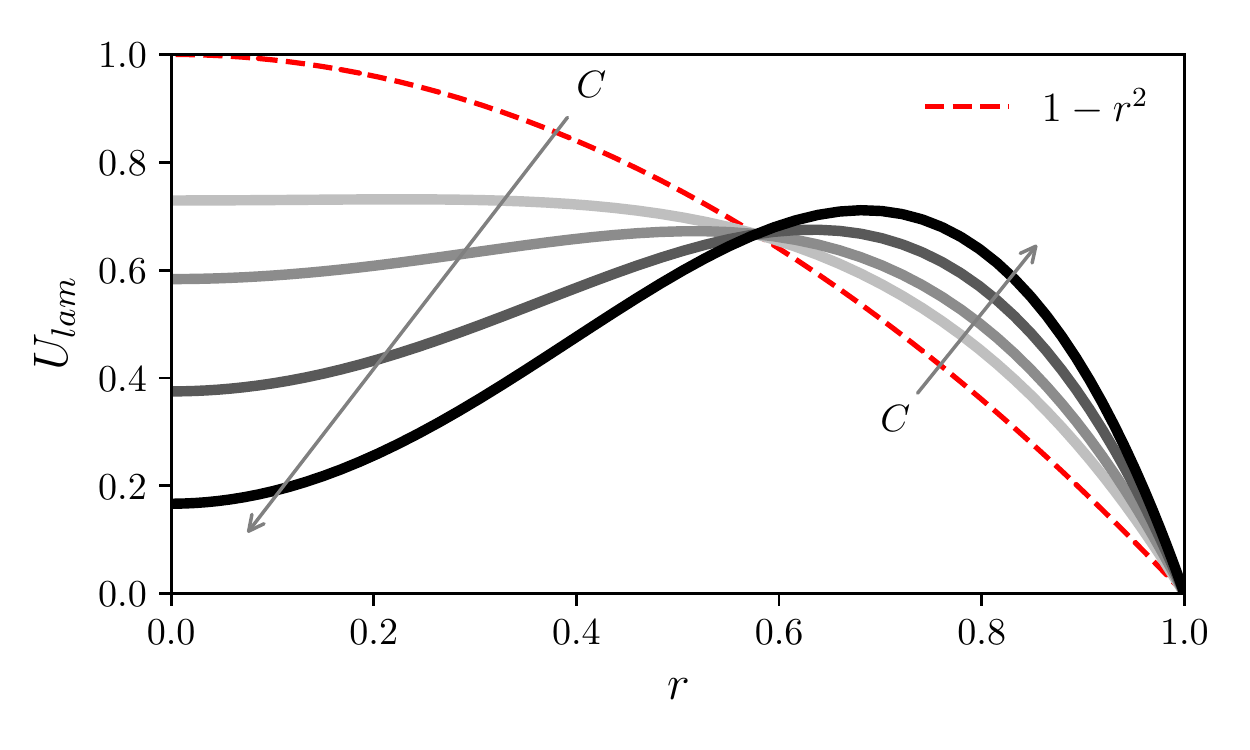}
\caption{\small (Left) Schematic of the flow configuration \Rtwo{A pipe section of length $L$ and radius $R$ is considered. The pipe is vertically aligned in the gravity field $\mathbf{g}$ and the fluid inside it is driven upwards by an externally applied pressure gradient and by buoyancy. The latter results from the lightening of the fluid close to the heated wall. We assume that the temperature at the wall $T_w$ remains constant in the pipe section}. (Right) Laminar \Rtwo{velocity profiles \eqref{Ub} for increasing values of $C$, as indicated by the arrows. Red dashed line: $C=0$ (isothermal profile)}. Light grey to black lines: $C=3$, 5, 7.5, 10.
  \label{fig-baseprof}}
\end{figure}

Consider a vertically aligned circular pipe of diameter $D$,
with the flow of fluid upwards.
%
%
We model a short pipe section of length $L$
 (figure \ref{fig-baseprof}(left)) 
 and let $\{\vec{u}(\vec{x},t),p(\vec{x},t),T(\vec{x},t) \}$
be the velocity, pressure and temperature fields, respectively.
%
%
The fluid has kinematic viscosity $\nu$, density $\rho$, volume expansion coefficient $\gamma$ and thermal diffusivity $\kappa$.
Under the Boussinesq approximation, density variations are ignored except where they appear in terms multiplied by the acceleration due to gravity, 
$g\,\hat{\vec{z}}$, leading to the governing equations
\beq
\nabla \cdot \vec{u} = 0 \, ,
\eeq 
\beq
\frac{\p \vec{u}}{\p t} + \vec{u}\cdot \bnabla \vec{u}= - 
\frac{1}{\rho}
\bnabla p 
+ \nu \,\nabla^2\vec{u} 
+ \frac{1}{\rho}\,(1+\beta)\,\mathrm{d}_zP\, \hat{\vec{z}}
+ \gamma\, g\, (T-T_{ref})\, \hat{\vec{z}} \, ,
\eeq
\beq \label{eq:dimTeqn}
\frac{\p T}{\p t} + \vec{u} \cdot \bnabla T = \kappa\, \nabla^2T 
 - \epsilon \, ,
\eeq
where $T_{ref}$ is a reference temperature
defined in the following subsection, and
$\mathrm{d}_zP$ is the pressure gradient for laminar flow with bulk
velocity $U_b$.  We suppose that $U_b$ is fixed, in which case 
$\beta(\vec{u})$ adjusts to maintain fixed bulk velocity.
\Rtwo{
We also suppose that the temperature of the wall $T_w$ and the bulk 
temperature $T_b$ are fixed. The latter is achieved by including 
a uniform heat sink}
$\epsilon(t)$ 
which adjusts to maintain the fixed bulk value 
$T_b$.  For such a flow, we can introduce axial periodicity, 
so that $\epsilon(t)$ may be considered
equivalent to the rate at which heat absorbed by the fluid 
would otherwise be carried out of the section of pipe\footnote{Spatial periodicity limits the domain over which wall friction is averaged, which can lead to unrealistic fluctuations (mean-square variations from the time average) in the bulk velocity. We therefore assume constant flux.}.

For laminar flow, the flow is purely axial so that
radial heat transport is purely conductive.  If  
$\epsilon_0$ is the heating rate for the laminar case, then
the observed quantity $\Nu:=\bar{\epsilon}/\epsilon_0$
is the Nusselt Number, where the overbar $\overline{(\bullet)}$ denotes time average.
\subsection{Non-dimensionalisation}
\label{sect:nondim}

\Rtwo{Given the temperature at the wall $T_w$ and the bulk temperature $T_b$,
we put $\Delta T=2(T_w-T_b)$ and 
take a reference temperature $T_{ref}=T_w-\Delta\,T=2T_b-T_w=T_c$, where 
$T_c$ is the centreline temperature 
for the case of laminar flow.
(The choice for $T_{ref}$ does not influence the flow, since the constant 
$\gamma\,g\,T_{ref}$ could be absorbed into the pressure gradient.)}
We
introduce the dimensionless temperature $\T =(T-T_c)/\Delta T$.
Let the pipe radius $R=D/2$
be the length scale and the isothermal laminar centreline velocity $2\,U_b$
be the velocity scale. \Rtwo{The corresponding time scale is thus $R/(2\,U_b)$.}
Hereafter, all variables are dimensionless except $\epsilon(t)$ which 
always appears in the dimensionless ratio $\epsilon/\epsilon_0$, 
i.e.\ the instantaneous Nusselt number.
Non-dimensionalising with these scales, for the temperature equation we find
\beq \label{eq:nondT}
\frac{\p \T}{\p t} + \uvec \cdot \nabla \T 
\,=\, \frac{\kappa}{2\,U_bR}\nabla^2 \T-\frac{\epsilon R}{2\,U_b\,\Delta T}\,.
\eeq
For the laminar case, $\T=\T_{lam} = r^2$, we find 
\beq \label{eq:Tscale}
   0 \,=\, \frac{\kappa}{2\,U_bR}\cdot 4 -\frac{\epsilon_0 R}{2\,U_b\,\Delta T}
   \quad\mbox{i.e.}\quad
\Delta T\,=\,\frac{\epsilon_0\,R^2}{4\,\kappa}\,.
\eeq
Plugging this $\Delta T$ back in to (\ref{eq:nondT}), we obtain 
the dimensionless temperature equation
\beq \label{temp:eq}
\frac{\p \T}{\p t} + \uvec \cdot \nabla \T \,=\, \frac{1}{Re\,Pr}\nabla^2 
\T - \frac{4}{Re\,Pr}\,\frac{\epsilon}{\epsilon_0}\,,
\eeq
where $Re:=U_bD/\nu$ is the Reynolds number and $Pr:=\nu/\kappa$ is the
Prandtl number. For the momentum equation we find
\beq
\frac{\p \uvec}{\p t} + \uvec \cdot \nabla \uvec= 
- \nabla p
 + \frac{1}{Re} \nabla^{2}\uvec
 + \frac{4}{Re} \,(1+\beta) \, \hat{\mathbf{z}}
+ \frac{\gamma \,g\, \Delta T \,R}{(2\,U_b)^2} \, \T \,
\hat{\mathbf{z}}
\eeq
The coefficient of the 
buoyancy term can be written
\beq
\frac{\gamma \,g\,\Delta T\, R}{4\,U_b^{2}} \,=\, 
\frac{1}{4}\,
\frac{\gamma \,g\,(T_w-T_b)\,D^3}{\nu^2} \,
\frac{\nu^2}{U_b^2\,D^2} 
\,=\, \frac{1}{4}\,Gr \, Re^{-2}\, ,
\eeq
where $Gr:=\gamma \,g\,(T_w-T_b)\,D^3/\nu^2  $  
is the Grashof number.
Although the Grashof number is in common use, from $Gr$
it is difficult to judge the magnitude of the
buoyancy force relative to the pressure gradient of the laminar flow 
for this particular configuration.
We therefore write the dimensionless 
momentum equation as
\beq
\label{eq:veleq}
\frac{\p \uvec}{\p t} + \uvec \cdot \nabla \uvec= - \nabla p + \frac{1}{Re} \nabla^{2}\uvec 
+\frac{4}{Re}(1+\beta+C\,\T)\,\hat{\mathbf{z}} \, ,
\eeq
where $C$ measures the buoyancy force 
relative to the force that drives the laminar isothermal shear flow,
\beq
C=\frac{Gr/(4\,Re^2)}{4/Re} := \frac{Gr}{16\,Re}.  
\eeq
The laminar velocity and laminar temperature profiles for this 
configuration are 
\beq
\label{Ub}
U_{lam}(r) = \left(1-r^2\right) 
+ C\left(\frac{1}{3}\,r^2-\frac{1}{4}\,r^4-\frac{1}{12}\right) \, ,
\qquad
\T_{lam}(r) = r^2 \, ,
\eeq
and the no-slip and fixed-temperature boundary conditions at $r=1$ are 
\beq
   \vec{u}=\vec{0},  \qquad \T=1 ,
\eeq
respectively, while periodic boundary conditions are applied in the streamwise direction. The laminar velocity profiles for different $C$ are shown in figure \ref{fig-baseprof}(right). 
The isothermal pipe flow is recovered for $C=0$ (no buoyancy force) and $Pr=0$ (temperature diffuses immediately), with the parabolic laminar  profile $U_0=1-r^2$.

For a statistically steady flow, Reynolds averaging is both time averaging 
 and cylindrical surface averaging, where the latter is denoted as
\beq
   \langle (\bullet) \rangle(r) := \frac{1}{2\pi L} \int_0^L\int_0^{2\pi} 
   (\bullet) \,\mathrm{d}\theta\, \mathrm{d}z\,.
\eeq
Turbulent fluctuations are calculated as deviations from the mean components of the flow, i.e. $\left\{ \mathbf{u}'(\mathbf{x},t), \Theta'(\mathbf{x},t)\right\}:= \left\{ \mathbf{u}(\mathbf{x},t), \Theta(\mathbf{x},t) \right\} - \{\langle \overline{\mathbf{u}}\rangle(r), \langle \overline{\Theta}\rangle(r)\}$.
\subsection{Numerics}
\label{sect:numerics}
Simulations were carried out using the \texttt{Openpipeflow} solver \citep{willis-2017},
modified to include timestepping of the temperature field and the buoyancy term in the momentum equation.
A variable $q(r,\theta, z)$ is discretised using 
a non-uniform grid  in the radial direction with points clustered near the wall 
and Fourier decompositions in the azimuthal and streamwise directions, namely
\beq \label{eq:discretisation}
q(r,\theta,z) = \sum_{k<|K|} \sum_{m<|M|} q_{km}(r_n)e^{i\alpha k z + m_p m \theta} \quad n=1,...,N
\eeq
where $\alpha=2\pi/L$ is the streamwise wavenumber and $m_p$
 determines the azimuthal periodicity ($m_p=1$ for no discrete rotational symmetry).
\Rone{Radial derivatives are evaluated using central finite differences with a nine-point stencil.} 
At $Re=5300$, in a $L=5D$ long pipe we use a spatial resolution of $(N \times M \times K) = (64 \times 96 \times 96)$,
which ensures a drop of at least 4 orders of magnitude in the spectra
and provides the correct value for the friction factor, as reported in the literature \citep{Eggels94}.
\Rone{Following the 3/2 dealiasing rule, variables are evaluated on an 
$N \times 3M \times 3K$ grid in physical space.}
\Rone{A second-order predictor-corrector scheme is employed for temporal discretisation, and a
fixed timestep of 0.01 is used.  This is sufficient to ensure
that the time discretisation error is no larger than the spatial discretisation
error (measured by the corrector and spectra respectively)
and corresponds to a CFL-number of approximately $0.2-0.25$.}  

%
\begin{figure}
\centering
\includegraphics[width=0.6\textwidth]{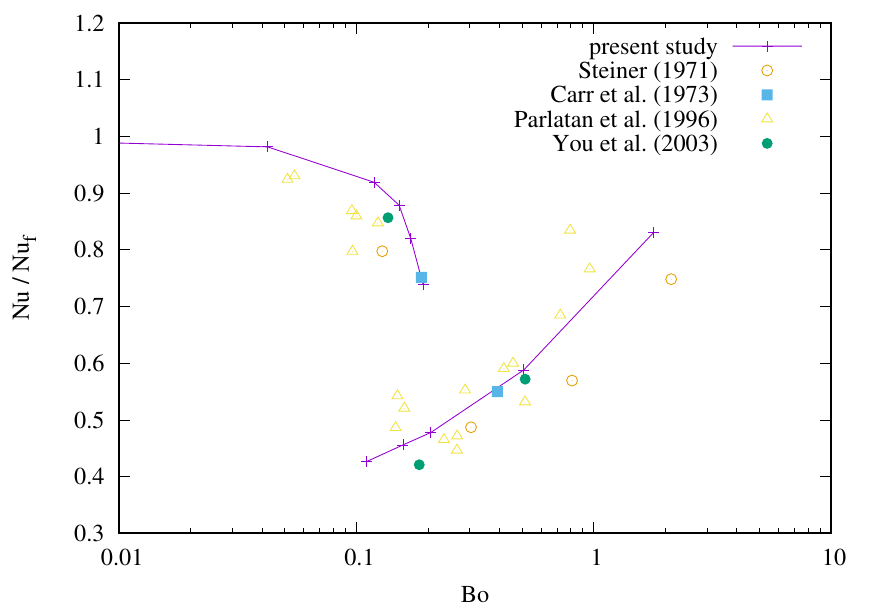}
\caption{\label{fig:BoNu}
\Rtwo{Change in $Nu$ flux, normalised by that for turbulent `forced convection' ($C \to 0$), as a function of 
$Bo\,=\,8\times 10^4 \, (8 \,Nu\, Gr) \,/\, (Re^{3.425} Pr^{0.8})$.  
Data from simulations at $Re=5300$, $Pr=0.7$ and various $Gr=16\,Re\,C$.
The upper and lower branches correspond to flow in shear-driven 
and convection-driven states respectively.}
}
\end{figure}
\Rtwo{Data for simulations for various $Gr=16\,Re\,C$ and constant $Re=5300,\,Pr=0.7$ 
are shown in figure \ref{fig:BoNu}.  There is good agreement with 
numerical data \citep{you-etal-2003} and experimental data
\citep{steiner1971reverse,carr1973velocity,parlatan1996buoyancy}.}\\


\subsection{Travelling wave solutions}
\label{sect:tw}
In order to apply dynamical systems theory, the discretised momentum and temperature equations are formulated as an autonomous dynamical system 
\citep{Visw07b,willis-etal-2013}:
\begin{equation}
\frac{\mathrm{d}\mathbf{X}}{\mathrm{d}t}=\mathbf{F}(\mathbf{X};\,\mathbf{p}),
\end{equation}
where $\mathbf{X}$ is the vector of dependent variables, here $\mathbf{X}=(\vec{u},\T)$, 
and $\mathbf{p}$ is the vector of parameters of the system, $\mathbf{p}=(Re,C)$.
 The simplest 
solution is an equilibrium, which satisfies $\mathbf{F}(\mathbf{X};\,\mathbf{p})=0$. 
For pipe flow, the only equilibrium solution is the laminar solution.
Travelling wave solutions satisfy
 $\mathbf{X}(\vec{x},t)=g(ct)\,\mathbf{X}(\vec{x},0)$, 
where here $g(l)$ applies a streamwise shift by $l$, and $c$ is the phase speed.
Travelling waves are also known as `relative' equilibrium solutions, as they are
steady in a co-moving frame.
They therefore satisfy
\begin{equation}\label{eq:GX}
\mathbf{G}(\mathbf{X}(0),l,T)=g(-l)\mathbf{X}(T)-\mathbf{X}(0)=\vec{0}\, ,
\end{equation}
 for some 
vector $(\mathbf{X},l,T)$, and hence can be calculated via a root solving method.
The most popular method at present is the Newton--Krylov method. 
(Note that in addition to (\ref{eq:GX}), 
two extra constraints are required to match the extra unknowns $l$, $T$; see \cite{Visw07b}.) 
Time-dependent periodic orbits may also be calculated by this method.  Typically 
periodic orbits originate via a Hopf bifurcation off a travelling wave, 
but are not discussed further in this work.
%

Stability of the solutions is calculated using the Arnoldi method 
to solve the eigenvalue problem 
\begin{equation} \label{eq:ex=Ax}
\mathrm{e}^{\sigma T}\,\mathbf{dX} = 
g(-l)\,(\mathbf{X}_0+\mathbf{dX})(T)-\mathbf{X}_0(0) \, ,
\end{equation}
where $\sigma$ is the growth rate and the operator on the right hand side is  
linearised about the travelling wave $\mathbf{X}_0$ 
 by taking $||\mathbf{dX}||\ll||\mathbf{X}_0||$.  
(Numerical performance is improved by replacing $\mathbf{X}_0(0)$ with 
$g(-l)\,\mathbf{X}_0(T)$ in (\ref{eq:ex=Ax}).)

The Newton-Krylov and Arnoldi solver,
 already available as a utility of \texttt{Openpipeflow} \citep{willis-2017}, were 
integrated with the time-stepping code described in \S \ref{sect:numerics} 
for heated pipe flow.

\section{Results and discussion}
\label{sec:results}

All results presented herein pertain to the case $Pr=0.7$ and constant volume flux.
This relatively low Prandtl number 
is a reasonable starting choice for the applications we are interested in, 
where most gasses have $Pr\approx 0.7$, e.g.\ CO$_2$. In large scale 
cooling applications using liquid metal, $Pr$ is much smaller.
Cases where $Pr>1$ (e.g.\ $Pr=7$ for water) are more expensive numerically due to a 
need for higher resolution for the temperature field.


\subsection{Direct Numerical Simulations}
\label{sec:DNS}

\begin{figure}
\centering
\includegraphics[width=0.6\textwidth]{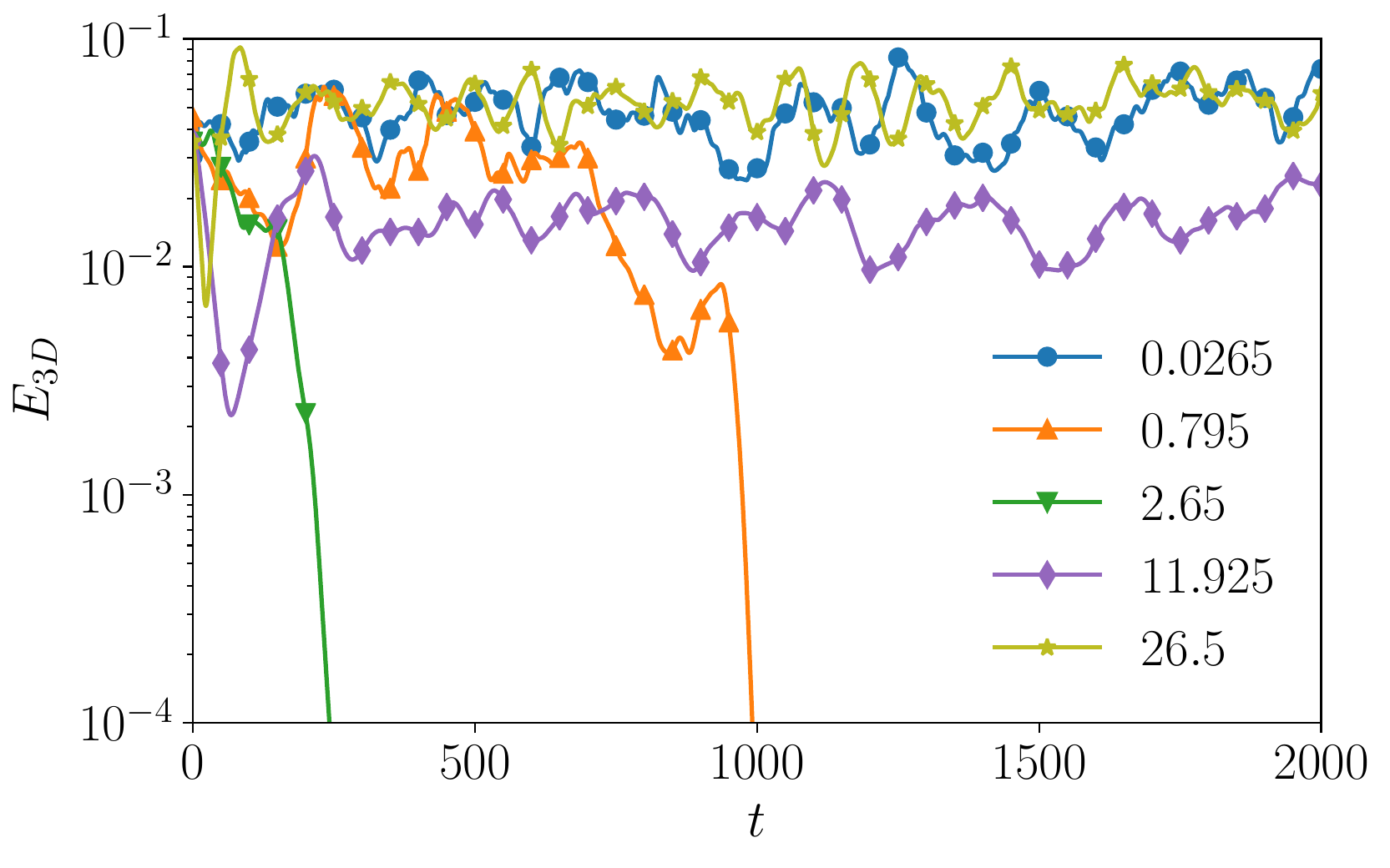}
\caption{\label{fig:sim2500}
   Energy of the streamwise-dependent component of the flow.
   $Re=2500$, $L=5D$, $Pr=0.7$
   for a range of $C$ \Rtwo{(values reported in the legend)}. Intermediate values of $C$ destabilise the turbulence,
   or even cause immediate relaminarisation.
}
\end{figure}

Simulations were performed in a pipe of length $L=5D$ for a range of Reynolds numbers to study the effect of the buoyancy parameter $C$.
%
Results are first shown for a relatively low Reynolds number, $Re=2500$.
Figure \ref{fig:sim2500} shows complete relaminarisation 
of transitional turbulence in response to the introduction of buoyancy 
for intermediate values of $C=\mathcal{O}(10^{-1} )-\mathcal{O}(1)$.
Relaminarisation events are revealed by monitoring the energy of the streamwise-dependent component of the flow, denoted $E_{3D}$, which shows a rapid decay when the flow
relaminarises, $E_{3D}\to 0$ and $\varepsilon(t)/\varepsilon_0\to 1$.
 At larger $C\ge\mathcal{O}(10)$, turbulent fluctuations are not completely suppressed.
Instead a convection-driven flow is set up, which becomes stronger as $C$ is increased.

\begin{figure}
\centering
\hspace{2cm}\includegraphics[width=0.65\textwidth]{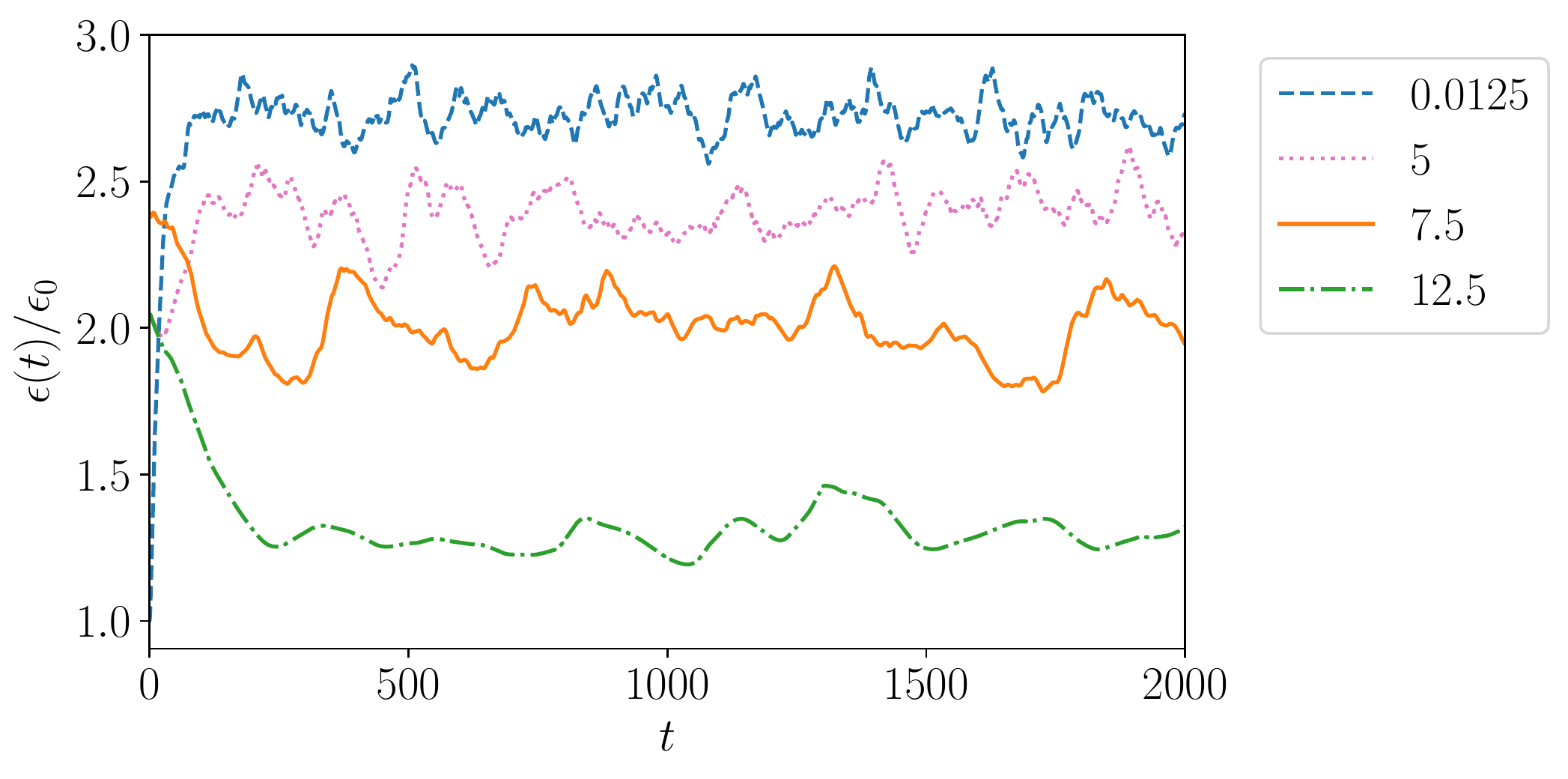}
\\
\includegraphics[width=1\textwidth]{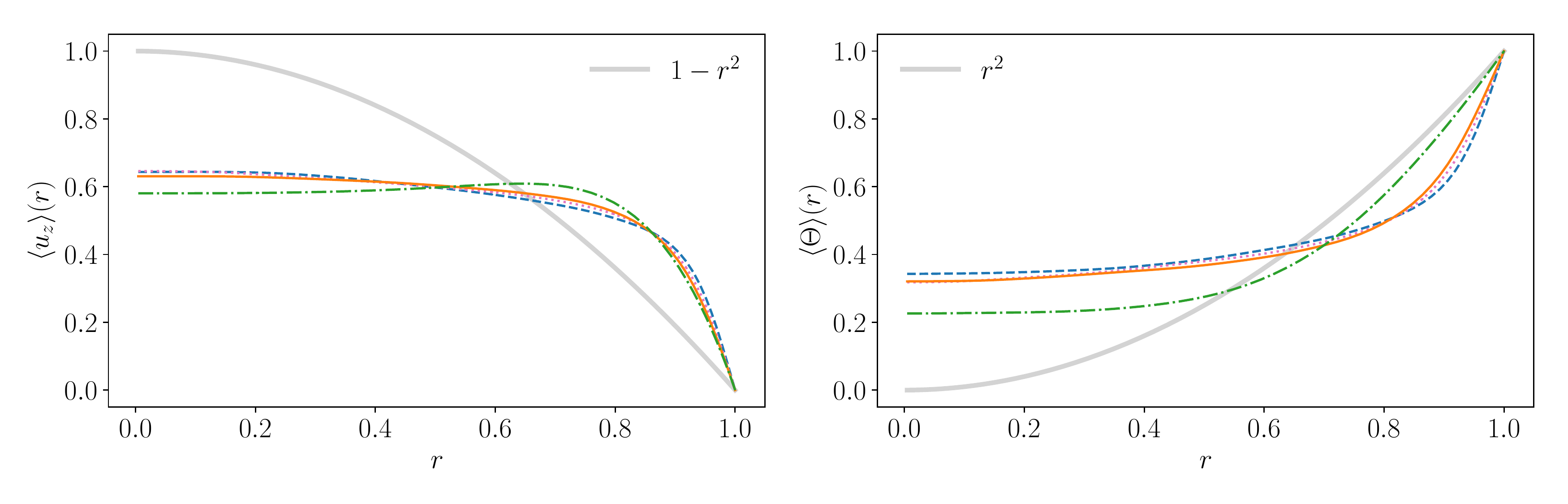}
\caption{\label{fig:sim5300}
$Re=5300$, 
$L=5\,D$, $Pr=0.7$, resolution 
$64\times96\times96$.
Top: Non-dimensional instantaneous heat flux, $\Nu=\epsilon/\epsilon_0$
\Rtwo{for different values of $C$, as indicated in the legend}.
Bottom: Snapshots of 
mean streamwise velocity $\langle u_z\rangle(r)$ and temperature $\langle \T \rangle(r)$ profiles
\Rtwo{at $t=1000$} 
 \Rtwo{for the same values of $C$ shown at the top. The thick light-grey lines correspond to the laminar profiles \eqref{Ub} with $C=0$.}
}
\end{figure}

At $Re=5300$ the effect of buoyancy is found to be slightly different --
turbulence is not observed to undergo complete 
relaminarisation, 
but instead transitions directly to a weak convection-driven state.
Figure \ref{fig:sim5300} shows simulations with 
 $C=O(1)-O(10)$.  The buoyancy causes suppression of the 
turbulence and therefore a drop in $\varepsilon(t)/\varepsilon_0$, so that 
the Nusselt
number $\Nu=\bar{\varepsilon}/\varepsilon_0$ reduces substantially. The corresponding velocity and temperature mean profiles, $\langle u_z\rangle(r)$ and $\langle \T \rangle(r)$,
 are shown in bottom graphs of figure \ref{fig:sim5300} together with the laminar profiles at $C=0$ for comparison. Cases where turbulence is suppressed exhibit a flattened base velocity profile.

\begin{figure}
\centering
\includegraphics[width=1\textwidth]{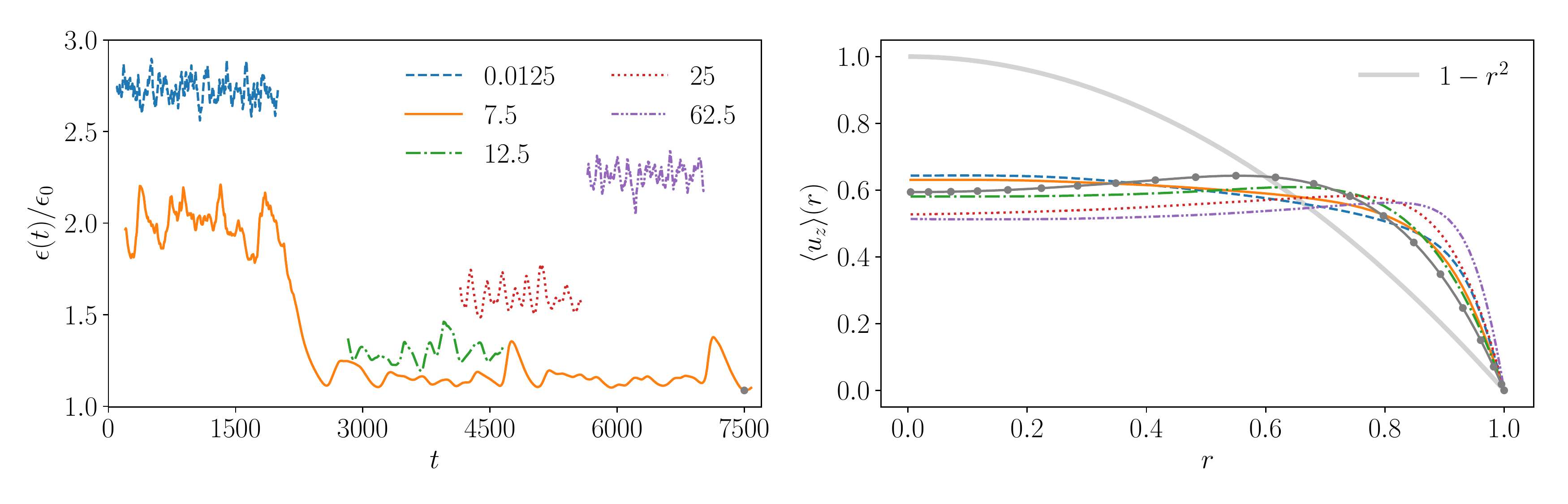}
\caption{\label{fig:sim5300b}
Parameters as in figure \ref{fig:sim5300} but for larger $C$ \Rtwo{(values reported in the legend)}. Left: Non-dimensional instantaneous heat flux, $\Nu=\epsilon/\epsilon_0$.
 \Rtwo{The initial transients ($t \approx 100-200$) are omitted for all trajectories and the curves corresponding to $C \ge 12.5$ are shifted in time by an arbitrary offset, for clarity only.}
 Right: Snapshots of the mean streamwise velocity profiles $\langle u_z\rangle(r)$ \Rtwo{for the same values of $C$ shown on the left.
  All the snapshots are taken at $t=1000$. For $C=7.5$ an additional snapshot (solid grey line with dots) is shown corresponding to $t=7500$ (marked with a grey dot on the corresponding trajectory on the left). The thick light-grey line on the right corresponds to the laminar streamwise velocity profile \eqref{Ub} with $C=0$.}}
\end{figure}

The case for $C=7.5$ is shown for longer time in figure 
\ref{fig:sim5300b}(left).
The shear-driven turbulent state is metastable only, and around $t \approx 2000$ 
turbulence is more suppressed as there is a switch to the more quiescent 
convection-driven state.  
As $C$ is increased further the buoyancy starts to drive a more turbulent 
convection-driven state.
For these cases the velocity profile is more `M-shaped' as seen in
figure \ref{fig:sim5300b}(right).

\begin{figure}
\centering
\includegraphics[width=1\textwidth]{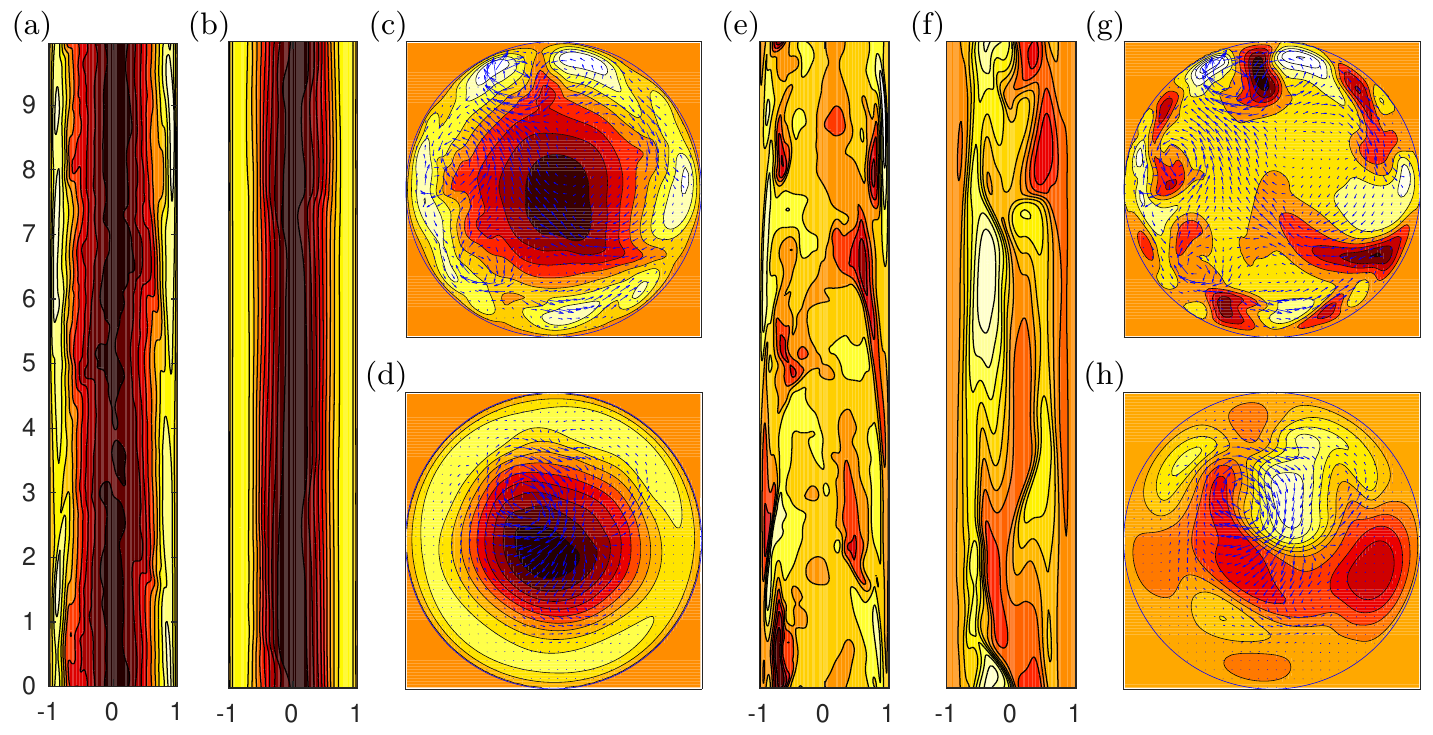}
\caption{\Rtwo{Isolevels of streamwise velocity perturbation for (a, c, e, g) the shear-driven turbulence and (b, d, f, h) the convective state at $Re=5300$,  $C=7.5$ and $t=1000$, $t=7500$, respectively. (The corresponding streamwise velocity profiles at these times were shown in \ref{fig:sim5300b}(right).)
Plots on the left (a--d) show deviations from the
isothermal laminar profile $U_0=1-r^2$, while plots on the right (e--h) 
show deviations from 
the mean profile $\langle u_z\rangle(r)$.
Dark/light regions correspond to slow/fast streaks. Ten contours are used between the maximum and minimum values, corresponding to (a--d) $u-U_0 \in [-0.4, 0.3]$, (e, g) $u' \in [-0.2, 0.1]$ and (f, h) $u'\in [-0.1, 0.08]$.
  The arrows in the  $r-\theta$ cross-sections (c, d, g, h) indicate the cross-sectional velocity components, multiplied by a factor of 2 for the shear turbulence (c, g) and 5 for the convective state (d, h), for visualisation reasons only. The  $r-\theta$ cross-sections (c, d, g, h) are taken at $z=0$ while the $r-z$ sections are taken at $\theta=\pi/2$.} }
\label{fig:Cvis}
\end{figure}

\Rtwo{The convective state at $Re=5300$ and  $C=7.5$ is visualised in figure \ref{fig:Cvis} together with the metastable shear-driven turbulent state.
When comparing the deviations from the isothermal laminar profile (a--d), both the shear and convective states show a deceleration in the core and acceleration close to the wall, with the convective states showing very smooth and almost $z-$ and $\theta-$independent contour levels.
Deviations from the mean profile (e--h), however, reveal that
the convective state has larger and more elongated flow structures
compared to the shear-driven turbulence.
In both types of visualisations it is clear that the small-scale turbulent eddies are strongly suppressed in the convection-driven flow.}

\begin{figure}
\centering
\includegraphics[width=0.9\textwidth]{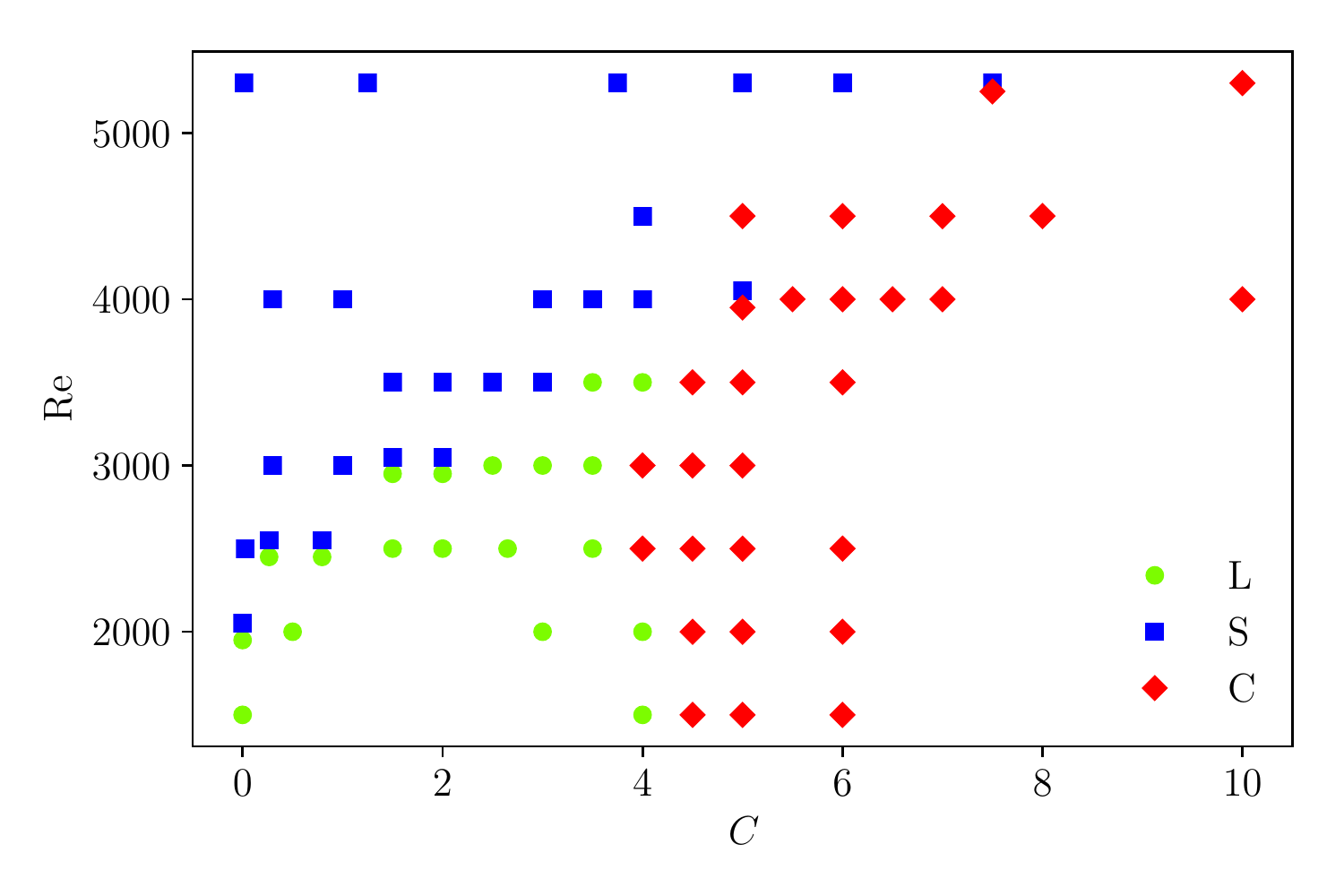}
\caption{ \label{fig:LTconv}
   Regions of laminar (L) flow, shear-driven (S) turbulence and 
   convection-driven (C) flow.  Points where multiple behaviours 
   are observed are marked with a slight offset in $Re$. \Rtwo{Simulations are initiated with a previously calculated shear-driven
state at similar $Re$, except for the region $Re\le2000$ and $C>3$ where
the shear-driven state decays immediately and hence simulations are started with a convection-driven state.}
}
\end{figure}

Figure \ref{fig:LTconv} shows the type of state seen in simulations, 
laminar flow (L), shear-driven turbulence (S) and convection-driven flow (C),
for a range of $Re$ and $C$.  
The initial condition for each simulation was a previously calculated shear-driven
state at similar $Re$. (This is except for $Re\le2000$ and $C>3$, where
it is clear that
\Rthree{the shear-driven state decays immediately, i.e. only the convective state could be supported,} and hence
the initial condition was of convection type).
For each simulation it is relatively easy to distinguish between the shear- and convection-type
flows, since the former shows far more chaotic time series and higher heat flux.  
The case for $C=7.5$ in figure \ref{fig:sim5300b}(left) shows this difference,
and also that multiple behaviours are possible at the same parameters for significant periods of time.
The shear-driven state is marked if observed for 
$\gtrsim 1000$ time units. 
\Rtwo{(It is stable or at least metastable with a long expected lifetime.)}
A relaminarisation is marked if the energy of the streamwise component
of the flow drops below $10^{-5}$. 
 %
Overall, figure \ref{fig:LTconv} indicates that as $C$ (or $Gr$) increases, a larger $Re$ is needed in order to drive shear turbulence, or, equivalently, as $Re$ increases, shear-driven states persist to larger $C$. For $C \ge 4$ simulations suggest that a convective instability kicks in, roughly independently of the Reynolds number over this range.
In between, it is possible to completely relaminarise flow up to $Re\approx 3500$,
but at larger $Re$ the progression is as in figure \ref{fig:sim5300b} -- 
from a shear-driven turbulent state to a weak convection-driven state, then to a more turbulent
convection-driven state as $C$ is increased.

In the following sections we determine whether the
boundaries of stability observed in figure \ref{fig:LTconv} are consistent with 
linear stability of the laminar flow, 
analysis of travelling wave solutions 
and the viewpoint of HHS.

\subsection{Linear stability analysis}
\label{sec:LS}

As the transition to shear-driven turbulence in isothermal flow occurs
in the absence of a linear instability, this section relates to the 
transition to convection-driven flow states, in particular
with respect to loss of stability of the \Rtwo{modified} 
laminar base profile \eqref{Ub} for non-zero $C$.
\Rone{Linear stability of mixed-convection pipe flow has been studied by \citet{yao-1987a, su-chung-2000}, where the model differs slightly in 
the boundary condition and form of the heat sink. Our figure
\ref{fig:BoNu} suggests these differences make little difference to
transition, however,
we check for consistency with the nonlinear results
of \S\ref{sec:DNS}.}

\Rtwo{As our code uses Fourier expansions in the periodic dimensions,
to calculate the eigenfunctions and stability of the base flow \eqref{Ub} we need simulate
only using a few Fourier modes.  The Arnoldi method is employed to 
accelerate convergence and to access eigenvalues beyond the leading one.
Linear stability analysis is performed for 
azimuthal wavenumbers $m=0,1,2$ and
two streamwise wavenumbers $\alpha=0.628$ and $\alpha=1.7$ 
(commensurate with the pipe lengths $L=5D$ and $L=1.85D$ used in our DNS study of \S \ref{sec:DNS} and in the travelling wave analysis of \S \ref{sec:N4L}).}

\begin{figure}
  \centering
        \includegraphics[width=0.9 \textwidth]{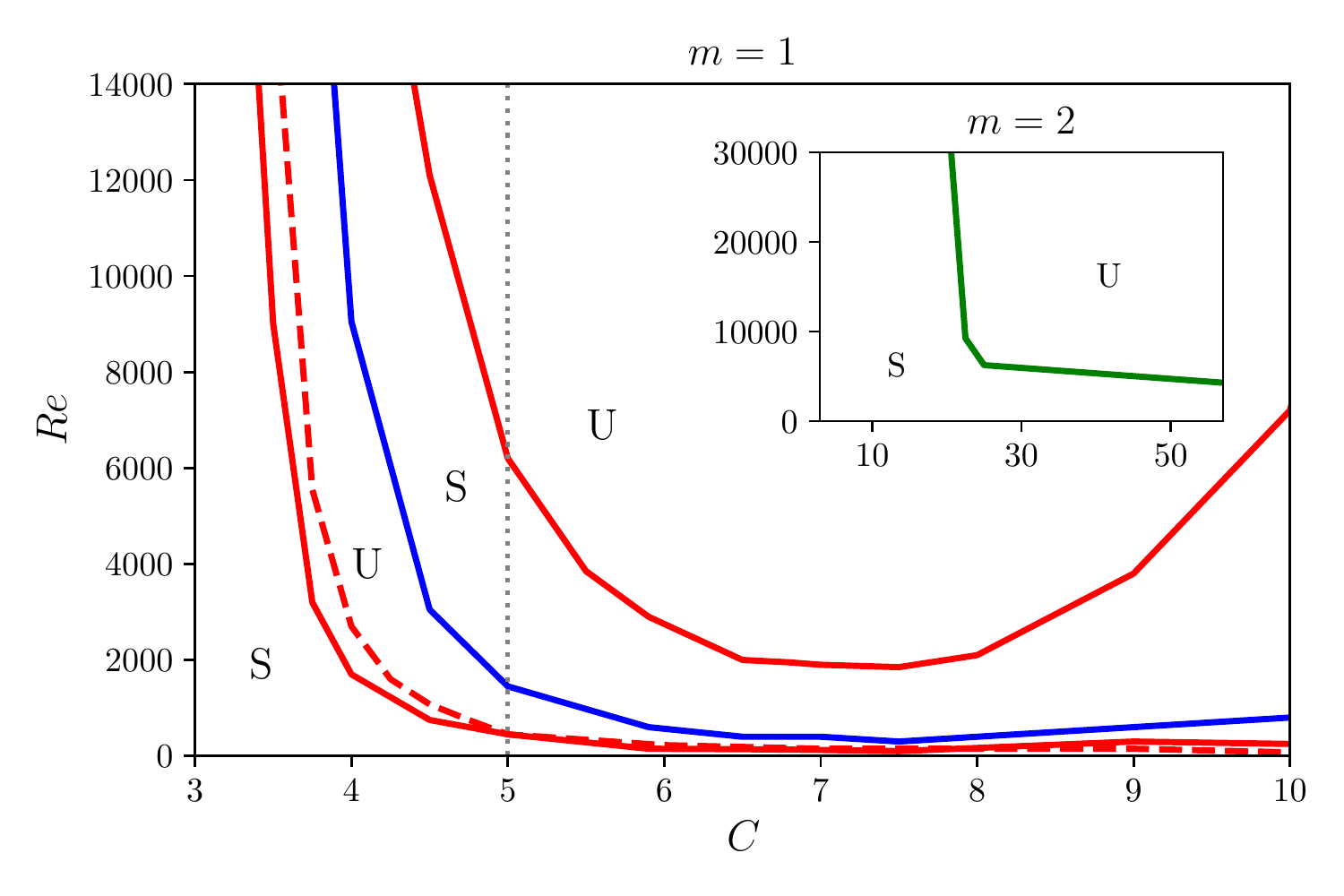}
  \caption{\small Linear stability analysis for $\alpha=1.7$, $k=1$ ($L=1.85D$) (solid lines). Main figure: $m=1$. Inset: $m=2$.
  \Rtwo{The axisymmetric mode is included in the $m=1$ analysis (i.e. $m=0$ and $\pm 1$), but instability of this mode is not observed. The first branch (for $m=1$) is also shown for the case $\alpha=0.628$ (dashed line).}
   \Rthree{The neutral curves delimit regions where the flow is linearly stable (S) or unstable (U).}  \Rtwo{
 The  dotted vertical line indicates the value of $C$ ($C=5$) at which the growth rate is shown in figure \ref{fig-heatedpipe-LS-C5} as a function of $Re$}.}
  \label{fig-heatedpipe-LS-1}
\end{figure}

The neutral curves, where the growth rate $\Re(\sigma)=0$, are shown in figure \ref{fig-heatedpipe-LS-1}. 
As expected \Rone{(and as also reported by \citet{yao-1987a, su-chung-2000})}, the first azimuthal mode is found to be the least stable,  
it corresponds to the spatially largest mode
\Rtwo{and is the only mode that can exhibit flow across the axis.
(The axisymmetric mode $m=0$ is included in the numerical calculations 
for stability of the $m=1$ mode, but we have not observed instability of 
$m=0$ type.)}
As shown in figure \ref{fig-heatedpipe-LS-1}, the $m=1$ mode exhibits
 a fairly complex dependence on $C$, \Rtwo{while it is only weakly affected by the axial wavenumber}.
 \Rtwo{Indeed, the first branch for $\alpha=0.628$ almost coincides with that for $\alpha=1.7$ and the other two branches (not shown) are slightly shifted to the right.}
Consistent with the linear stability of isothermal pipe flow, 
the critical Reynolds number approaches infinity as $C\to 0$ for any $m$.

Consistent with the appearance of the convective state found in simulation (figure \ref{fig:LTconv}), 
at $C \approx 4$ a linear instability appears, roughly independent of $Re$ for most of the
range considered.
The corresponding laminar profiles for $C = 3 - 10$ are shown in figure \ref{fig-baseprof}(right). 
For $C> 4$ the profiles present an ``M-shape'' (independent of $Re$, see \eqref{Ub}),
which becomes increasingly more pronounced as $C$ increases.  The difference at the centreline
is more than 80\% for $C=10$. The profile at $C=3$ is flatter than the parabolic (isothermal) profile, with a centreline difference of almost 30\%, but does not have any inflection point. \Rone{Therefore, in agreement with previous experimental and theoretical studies \citep{scheele-hanratty-1962, yao-1987a, su-chung-2000}, our analysis suggests that the linear instability of buoyancy-assisted pipe flow is linked to the inflectional velocity profiles ocurring at sufficiently large heating and it is almost independent of $Re$.}

\begin{figure}
  \centering
        \includegraphics[width=0.9 \textwidth]{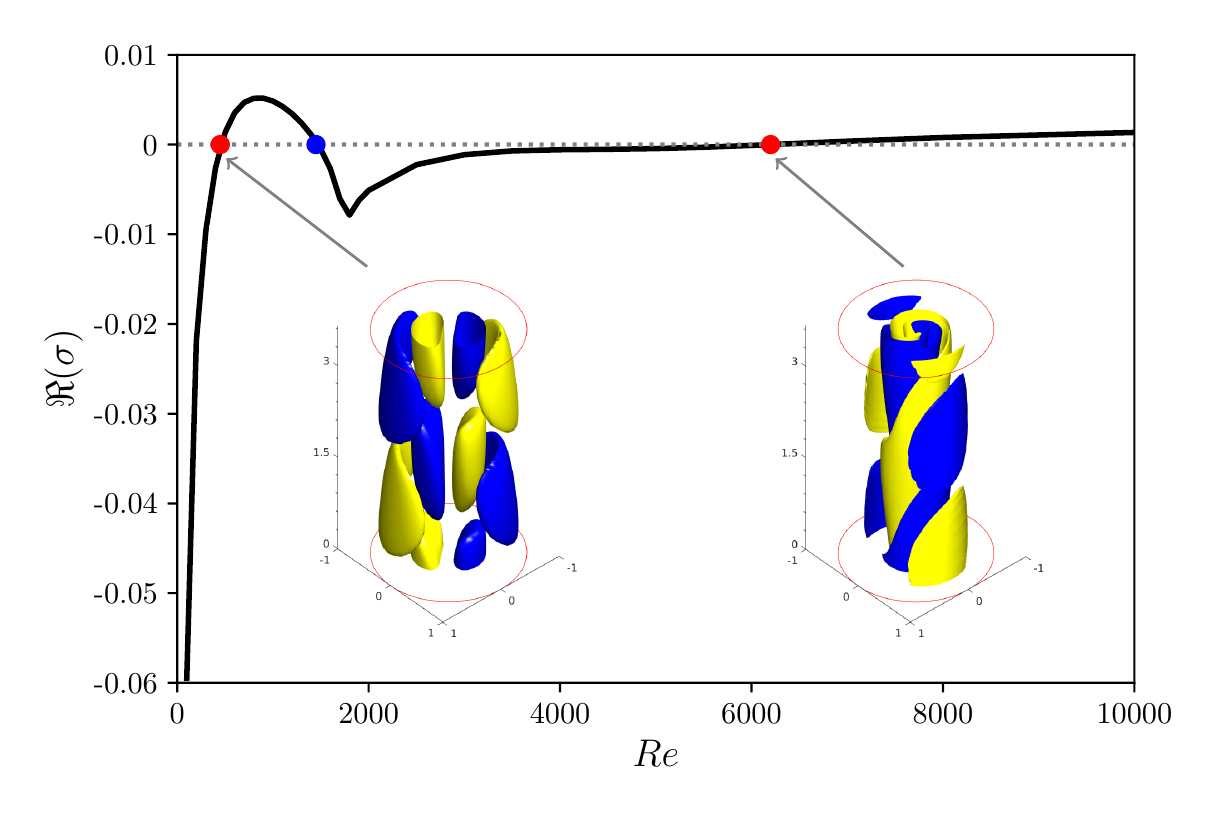}
  \caption{\small Growth rate vs Reynolds number from linear stability analysis at $\alpha=1.7$, $k=1$ ($L=1.85D$), $m=1$ and $C=5$ 
  \Rtwo{(corresponding to the dotted vertical line in figure \ref{fig-heatedpipe-LS-1}).} 
  Insets: streamwise vorticity (blue/yellow are 30\% of the min/max value) close to the two neutral points ($Re \approx 400$ and 6200).}
  \label{fig-heatedpipe-LS-C5}
\end{figure}

Figure \ref{fig-heatedpipe-LS-1} also shows that, for $C \gtrsim 4$, a region of restabilisation is observed as $Re$ is
increased. This is also evidenced in figure \ref{fig-heatedpipe-LS-C5}, which shows
a region of negative $\Re(\sigma)$ for $1450<Re<6200$ at $C=5$. 
Isosurfaces of streamwise vorticity for the eigenfunctions corresponding to the two neutral points where $\Re(\sigma)$ becomes positive ($Re \approx 400$ and 6200) are also shown in the insets of figure \ref{fig-heatedpipe-LS-C5}. For the larger Reynolds number, $Re  \approx 6200$, the eigenfunction looks like it is spiralling in the centre and resembles the ``spiral'' solution found by \cite{senoo-deguchi-nagata-2012}, although their visualised solutions are nonlinear.

\subsection{Continuation from $\mathrm{TW_{N4L}}$}
\label{sec:N4L}
To better understand the effect of buoyancy, we perform a nonlinear 
analysis,
 starting from a known TW in isothermal pipe flow ($C=Gr=0$) and continuing the solution
 to larger values.
A vast repertoire of TWs has now been compiled in isothermal pipe flows \citep{WFSBC15}. 
For our purpose
 we decided to focus on a fundamental solution, \Rone{labelled $\text{TW}_{\text{N4L}}$ \citep{Pringle09},
which is highly-symmetric (satisfying both shift-reflect and shift-rotate symmetries)
and characterised by
 relatively smooth continuation branches
 in order to aid the numerical continuation}.
\Rtwo{ In \cite{willis-etal-2013}, the lower branch of this solution was
found to lie on the boundary between the laminar state and turbulence 
in a `minimal flow unit'.  Localised solutions bifurcate off this class 
of solutions \citep{ChWiKe14} and are found to mediate transition in
extended domains \citep{AvMeRoHo13,budanur2017heteroclinic}}.

Following \cite{WiShCv15} we start with the `minimal flow unit' 
at Reynolds number $Re=2500$ with domain
 $(r, \theta, z) = [0,\,1] \times [0, \, \pi/2] \times [0, 2\pi/1.7] $, i.e. $m_p=4$ and $\alpha =1.7$ in (\ref{eq:discretisation}).
For isothermal flow ($C=Gr=0$), the phase speed of $\text{TW}_{\text{N4L}}$
is  $c=0.61925$. 
The isothermal TW
was first reconverged at $Pr=0.7$ using the Newton solver. A parametric continuation in 
$C$ to non-zero values was then performed (figure \ref{fig-heatedpipe-N4Lfixedflux-1})
for fixed $Re$, $Pr$ and $\alpha$. We were able to continue the isothermal solution
from $C=0$ around positive $C$ and find that it connects with the upper branch at $C=0$, then beyond to $C\approx -40$.  
(Negative $C$ corresponds to a downward cooled flow; see Appendix \ref{sect:link}).
As a check, we verified that the values of $c=0.52575$ and $Nu=2.378$ at $C=0$ on the upper branch, as well as the mean profiles, matched those of the previously known
upper-branch isothermal solution $\text{TW}_{\text{N4U}}$ with $Pr=0.7$.

In figure \ref{fig-heatedpipe-N4Lfixedflux-1}(right) it is seen that from $C=0$ to $C=6$ the 
Nusselt number $\mbox{\it Nu}$ increases by approx 0.75.  By comparison, along the upper
branch, over the large range $C=6$ to $C=-40$, it increases by only a further 1.25.
Relatively speaking, the lower branch is rapidly pushed back towards the upper branch
over the increase in $C$ and is suppressed altogether for $C>7.5$. 
%
\begin{figure}
  \centering
        \includegraphics[width=1 \textwidth]{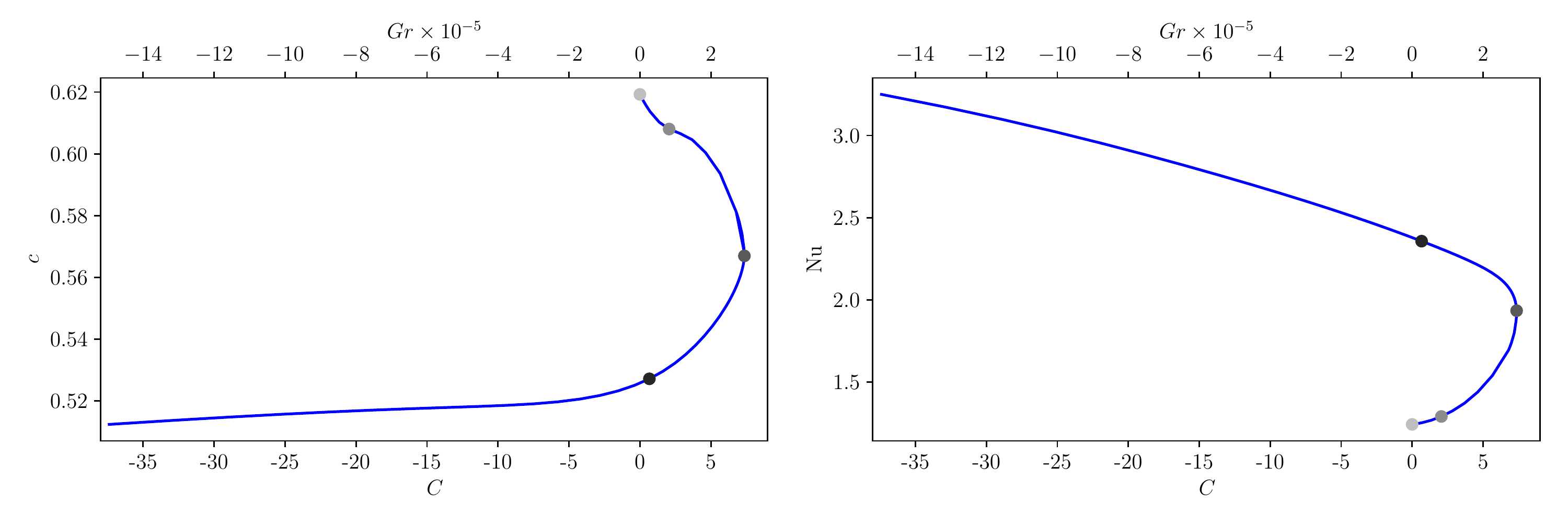}
  \caption{\small Continuation in $C$ (or $Gr$) from N4L at $Re=2500$. (a) Phase speed $c$ vs $C$ (or $Gr$), (b) $\mbox{Nu}$ vs $C$ (or $Gr$). Filled circles indicate the points along the continuation at which the mean streamwise velocity and temperature profiles are shown in figure \ref{fig-heatedpipe-N4Lfixedflux-2}.
}
  \label{fig-heatedpipe-N4Lfixedflux-1}
\end{figure}
The mean velocity and temperature profiles at different points along the continuation are  shown in figure \ref{fig-heatedpipe-N4Lfixedflux-2}.  
Observe that the profile in the near-wall region, where rolls and streaks occur, 
is similar at the saddle-node point (SN) to that of the isothermal upper branch (UB) solution.
\begin{figure}
  \centering
        \includegraphics[width=1 \textwidth]{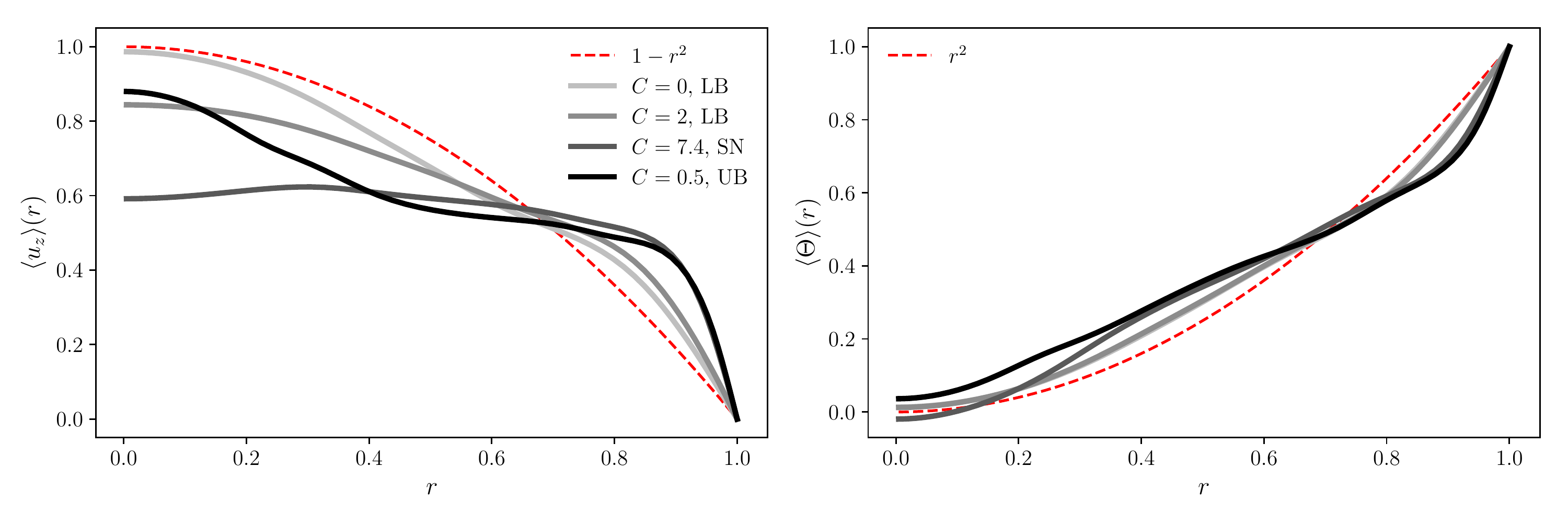}
  \caption{\small Mean streamwise velocity (left) and temperature (right) profiles at the points along the continuation from N4L ($Re=2500$) marked in figure \ref{fig-heatedpipe-N4Lfixedflux-1} \Rtwo{(SN: saddle node, LB/UB: lower/upper branch)}. The temperature profiles for $C=0$ and $C=2$ on the lower branch are indistinguishable.}
  \label{fig-heatedpipe-N4Lfixedflux-2}
\end{figure}
Figure \ref{fig-heatedpipe-N4Lfixedflux-3}(left) shows these rolls (arrows) and streaks (contours) 
in cross sections of the velocity perturbation at the saddle-node point. The corresponding temperature perturbation field (`thermal streaks') is shown on the right.
\Rtwo{Similar to its isothermal counterpart, the travelling wave is characterised by fast streaks located near the pipe wall and slow streaks in the interior. The core shows a strongly decelerated region relative to the laminar (isothermal) profile and thus the profile must become steeper at the wall to preserve the mass-flux.
The difference from the isothermal $\mathrm{TW_{N4L}}$, however, is less marked in the near-wall region than it is in the core.}
\begin{figure}
  \centering
          \subfloat{\includegraphics[width=0.45\textwidth]{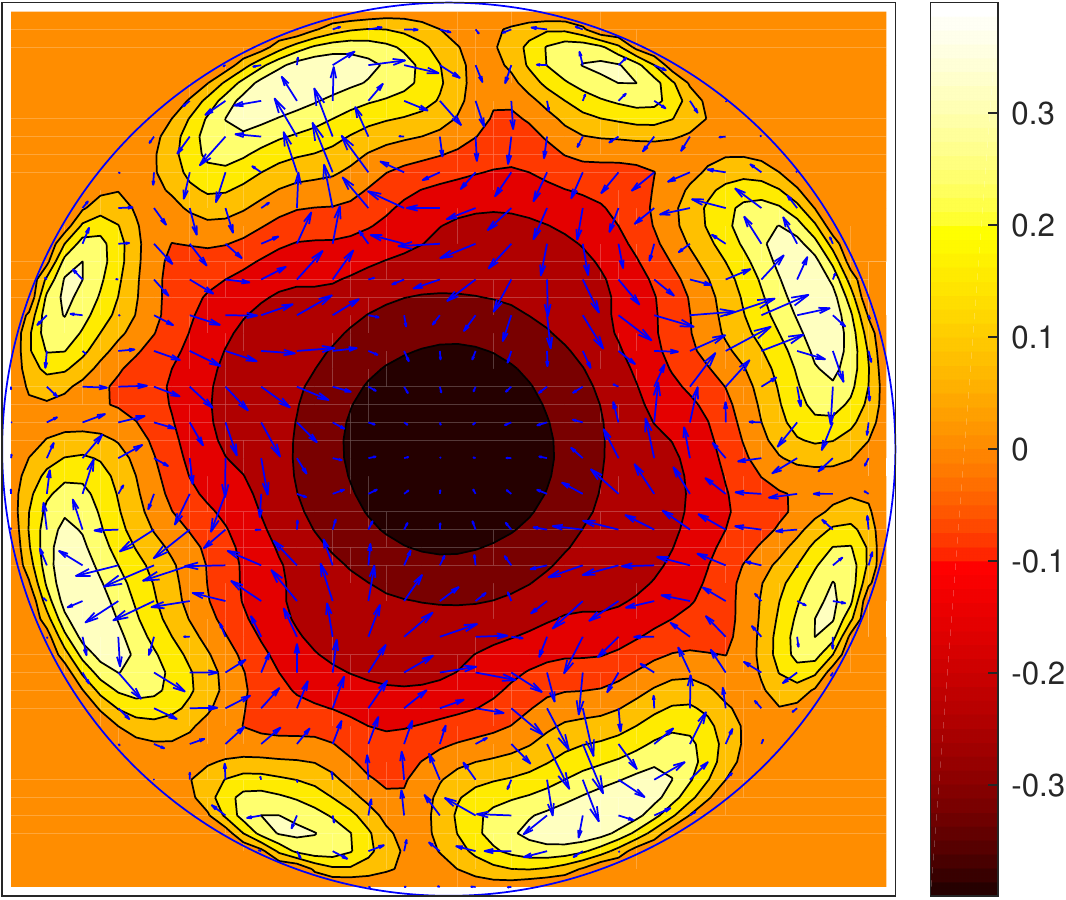}}\quad \quad
        \subfloat{\includegraphics[width=0.45 \textwidth]{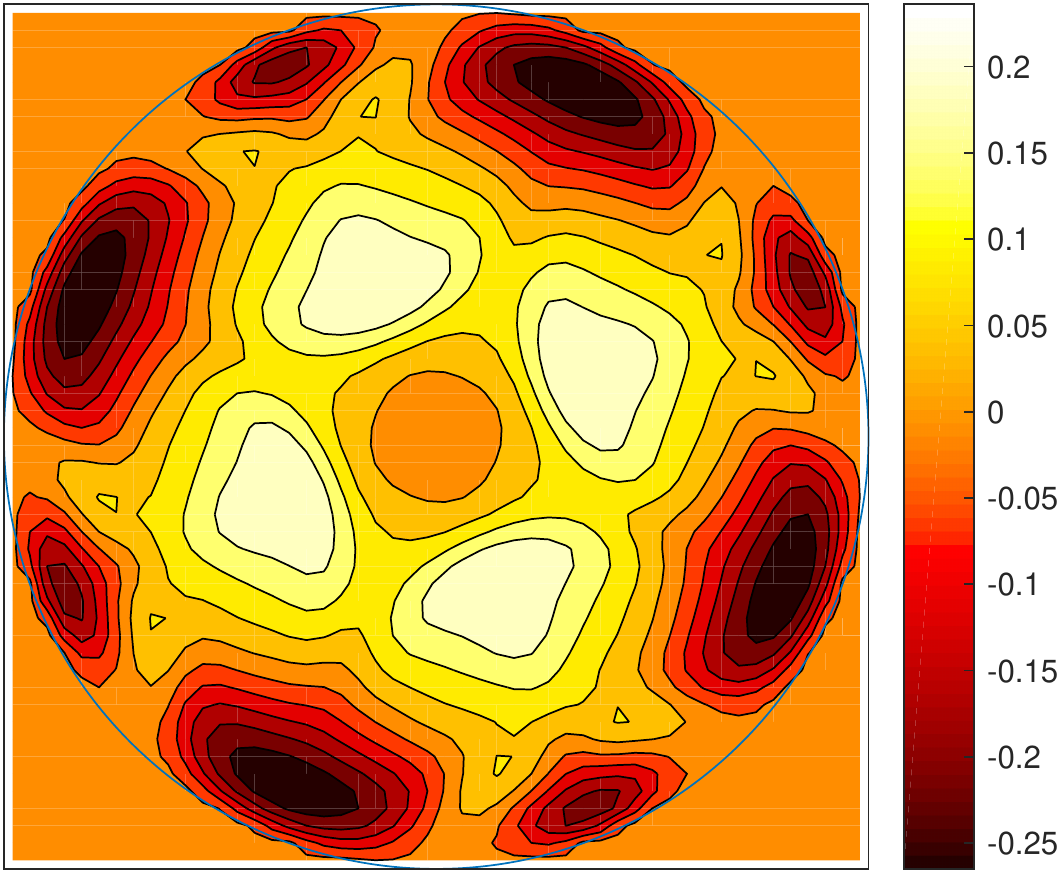}}\\
  \caption{\small Cross sections of streamwise velocity (left) and temperature (right) perturbations (deviations from the isothermal laminar flow) for the N4L travelling wave at $Re=2500$ and $C=7.4$ (saddle node). Ten contours are used between the maximum and minimum. The arrows in the left graph indicate the cross-sectional velocities}
  \label{fig-heatedpipe-N4Lfixedflux-3}
\end{figure}

Continuations were also performed at $Re=2000$ and $3000$, after reconverging the isothermal $\text{TW}_{\text{N4L}}$ at these Reynolds numbers. Results are shown in figure \ref{fig-HeatedPipe-N4Lfixedflux-NuvsGr-Re}.
\begin{figure}
  \centering
  \includegraphics[width=0.7 \textwidth]{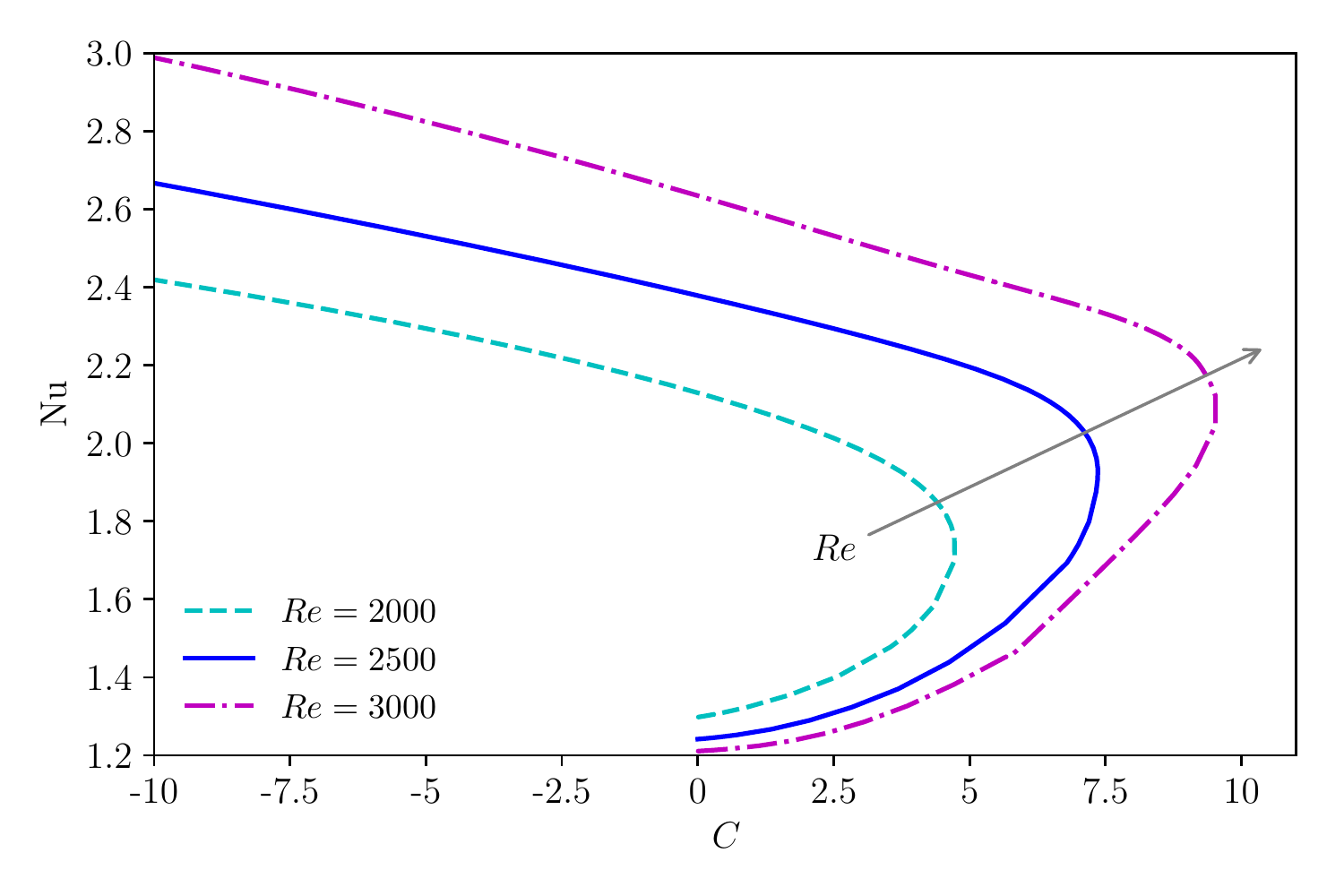}
  \caption{\small Continuation in $C$ from N4L \Rtwo{for increasing values of $Re$. The curve for $Re=2500$ is the same as that shown in figure \ref{fig-heatedpipe-N4Lfixedflux-1}(right)}.}
  \label{fig-HeatedPipe-N4Lfixedflux-NuvsGr-Re}
\end{figure}
The TW survives to larger $C$ as the Reynolds number increases (the saddle-node point of each curve moves to larger $C$ as $Re$ increases). This is consistent with the shear turbulence region in figure \ref{fig:LTconv} persisting to larger $C$ as $Re$ is increased. The saddle-node bifurcations at each $Re$ occur at much larger values of $C$ than those at which suppression of turbulence was observed in the DNS. For example, at $Re=2500$ the saddle-node bifurcation occurs at $C\approx 7.5$, while in figure \ref{fig:LTconv} shear-turbulence survives only for 
$C \lesssim 1$.
This is not so surprising, considering that in isothermal pipe flows 
\Rtwo{ the lowest $Re$ at which the N4L travelling-wave solution is found, i.e. $Re=1290$ \citep{Pringle09}},
is much below the commonly observed value for transition in experiments ($Re \approx 1800 - 2300$). Furthermore, it should be taken into account that only one TW solution is analysed here -- it cannot capture the entire phenomenon of turbulence suppression in a heated pipe flow, although is found to capture some of the fundamental 
characteristics.
 
\begin{figure}
  \centering
  \includegraphics[width=1 \textwidth]{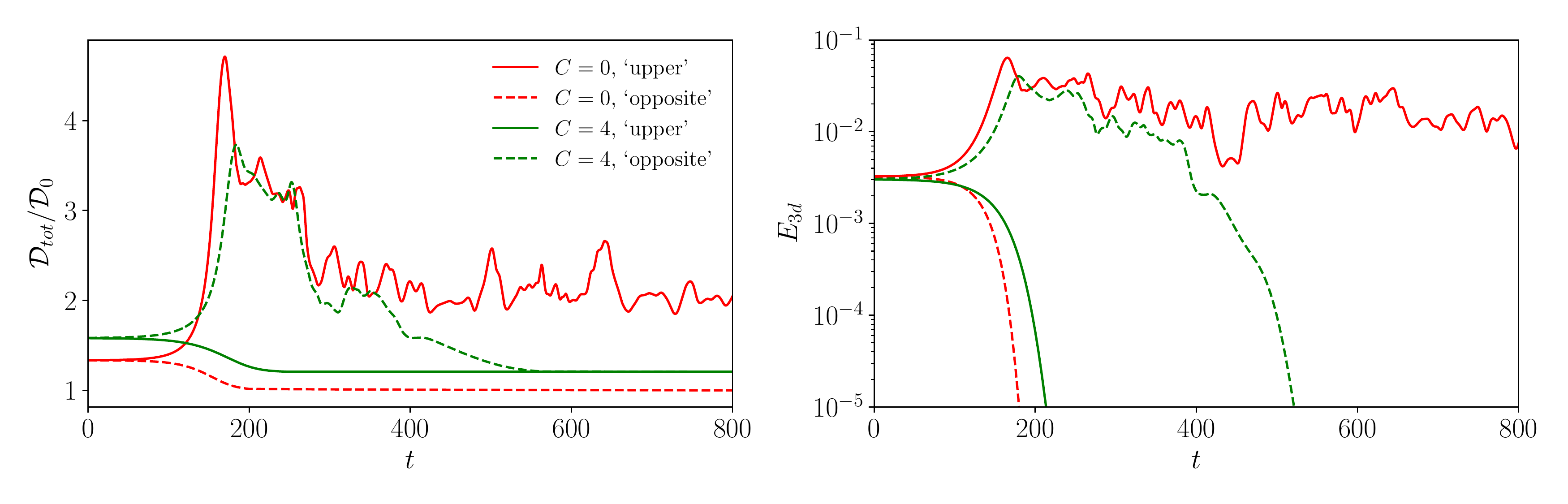}
  \caption{\small Times series of (left) total dissipation $\mathcal{D}_{tot}$ (normalised by the laminar isothermal value $\mathcal{D}_0=2\pi L_z |-2|=4\pi L_z$) and (right) energy of the streamwise-dependent modes $E_{3d}$ for simulations started from the lower-branch TW solutions at $Re=3000$, $\alpha=1.7$ with $C=0$ and $C=4$. The TW is perturbed by adding $\mp \,0.001\,(\mathbf{w}_1 +0.01 \mathbf{w}_2)$ (denoted as `upper' and `opposite' directions) where $\mathbf{w_1}$ and $\mathbf{w_2}$ are the first (leading) and second eigenvectors. Shooting in the `upper' direction leads to turbulence for $C=0$, while the flow goes back to laminar when perturbed in the opposite direction. For $C=4$ both directions end up at the laminar point.}
  \label{fig-HeatedPipe-TWdynamics}
\end{figure}

Figure \ref{fig-HeatedPipe-TWdynamics} shows that,
while the lower branch solution for $Re=3000$ is on the edge of an attractor for shear-driven
turbulence at $C=0$, this is no longer the case for $C=4$.  Shear-driven
turbulence does not survive in the heated case, although
shooting in the upper direction for $C=4$ does still produce a short turbulent transient.
In particular, large
amplification of the initial disturbance still occurs in the heated case,
but the self-sustaining mechanism appears to be disrupted.
 
To summarise this section, we have observed that 
\Rtwo{a known TW solution of the isothermal pipe flow is suppressed by buoyancy and that it is}
 connected to the transition {\em to} turbulence.  The observations 
are consistent with destabilisation of the shear-driven turbulent state,
but at this stage another approach is required to forge an approximate
quantitative link with the transition {\em from} turbulence.

\subsection{Calculation of the apparent Reynolds number of HHS}
\label{sec:HHS}
\Rtwo{In \S\ref{sect:HHSoverview}, where we gave a brief overview of HHS,
the (isothermal) 
equivalent pressure gradient flow (EPG flow) was identified as 
a useful reference case for heated flows.  To calculate
the apparent Reynolds number of the 
EPG reference flow, one must determine the contribution to the mass
flux from the buoyancy force that would have been induced in
a fixed pressure-gradient flow.
}
Here we summarise the key points of the analysis of HHS
and apply it to a selected example case from our data. (The interested reader is referred to sections 3.3 and 3.5 of HHS for a detailed derivation.) In the following section
we relate HHS analysis to the 
phase diagram determined from the simulations of \S \ref{sec:DNS}.

The analysis starts by decomposing the body-force influenced flow (i.e. the total flow) into a pressure-driven flow of equivalent pressure gradient (the EPG reference flow) and a perturbation flow due to the body force, \Rtwo{
\beq
\mathbf{u}(\mathbf{x},t) = \mathbf{u}^{\dagger}(\mathbf{x},t) + \mathbf{u}^f(\mathbf{x},t)
\label{eq:ssh-decomp}
\eeq
where the superscripts ${\dagger}$ and $f$ denote the EPG and the body-force perturbation driven flows, respectively.}
In contrast to the conventional view, HHS observe that 
\Rtwo{adding a non-uniform (radially-dependent) streamwise body force }
 to a flow initially driven only by a pressure gradient does not alter its turbulent mixing characteristics and in particular the turbulent viscosity remains approximately the same.
From this point of view, the body-force influenced flow behaves in the same way as the EPG flow 
and relaminarisation occurs when the Reynolds number $Re_{app}$ of this `apparent' flow drops below a certain threshold 
where turbulence cannot be sustained any more.
\Rtwo{Given the difficulties discussed in \S \ref{sec:intro} to uniquely define a critical Reynolds number for transition, we decided to follow HHS and select a nominal value of 2300, as quoted in many engineering textbooks \citep[see e.g.][]{white-book-1979}. }
By writing the bulk velocity $U_b$ of the EPG flow 
 as the difference between that of the total flow and of the body-force perturbation driven flow,
  i.e. $U_b^{\dagger}=0.5-U_b^{f}$, the above relaminarisation criterion can be expressed as
\begin{equation}
Re_{app} := Re\,\left(1-2\,U_b^f\right) < 2300\,.
\label{eq:hss-3.14}
\end{equation}

To determine $U_b^f$, the following expression was derived 
by integrating three times the Reynolds-averaged $z$-momentum equation of the body-forced perturbation flow:
\begin{equation}
U_b^f:=Re\left[ \underbrace{\frac{1}{2}\int_0^1 (1-r^2)f(r)\,r\mathrm{d}r}_{\mathcal{I}_1} + 
\underbrace{\int_0^1 r\mathscr{R}_{uv}^f(r) \,r\mathrm{d}r}_{\mathcal{I}_2}\right]
\label{eq:hss-3.12}
\end{equation}
where $\mathscr{R}_{uv}^f(r):=\langle\overline{(u_z'u_r')^f}\rangle$ is the Reynolds shear stress due to the perturbation flow induced by the body force $f(r)$.
The first integral of \eqref{eq:hss-3.12}, \Rtwo{ $\mathcal{I}_1:=\frac{1}{2}\int_0^1 (1-r^2)f(r)\,r\mathrm{d}r$}, represents the direct contribution of the body force (which is assisting the flow), while the second integral, \Rtwo{$\mathcal{I}_2:=\int_0^1 r\mathscr{R}_{uv}^f(r) \,r\mathrm{d}r$}, corresponds to the turbulent contribution related to the body-force perturbed flow.
The Reynolds stress term  $\mathscr{R}_{uv}^f$ 
of the body-force perturbed flow is related to that of the total ($\mathscr{R}_{uv}$) and EPG ($\mathscr{R}_{uv}^{\dagger}$)  flows by using the decomposition \Rtwo{\eqref{eq:ssh-decomp}} and
 is \Rtwo{approximated} by introducing the eddy viscosity concept,
\Rtwo{
\begin{equation}
\mathscr{R}_{uv}^f(r) = \mathscr{R}_{uv}(r)-\mathscr{R}_{uv}^{\dagger}(r) = \frac{\nu_t}{Re}\frac{\mathrm{d} \mathscr{U}_z}{\mathrm{d}r} - \frac{\nu_t^{\dagger}}{Re}\frac{\mathrm{d} \mathscr{U}_z^{\dagger}}{\mathrm{d}r}\,,
\label{eq-ssh-reynstress}
\end{equation}
where $\mathscr{U}_{z}(r):= \langle\overline{({u}_z)}\rangle$, $\mathscr{U}_{z}^{\dagger}(r):= \langle\overline{({u}_z)^{\dagger}}\rangle$ and $\nu_t$  and $\nu_{t}^{\dagger}$ are the eddy viscosities of the total and EPG flows, respectively.}
Under the assumption that  $\nu_t = \nu_{t}^{\dagger}$, we obtain
\begin{equation}
\mathscr{R}_{uv}^f(r) = -\frac{\nu_t^{\dagger}}{Re}\frac{\mathrm{d} \mathscr{U}_z^f}{\mathrm{d}r}\,,
\label{eq-3.7-HHS}
\end{equation}
where the perturbation flow $\mathscr{U}_{z}^f(r):= \langle\overline{({u}_z)^{f}}\rangle$ due to the imposed body force is obtained by integrating the Reynolds-averaged $z$-momentum equation 
\begin{equation}
0 = \frac{1}{r}\frac{\mathrm{d}}{\mathrm{d} r}\left[\frac{r}{Re}\left((1+\nu_t^{\dagger})\frac{\mathrm{d} \mathscr{U}_{z}^{f}}{\mathrm{d} r} \right) \right] +f \, ,
\label{eq-3.6-HHS}
\end{equation}
provided that the EPG flow (and hence $\nu_t^{\dagger}$) is known.
Equations \eqref{eq-3.7-HHS} and \eqref{eq-3.6-HHS} correspond to equations (3.6) and (3.7) of HHS and the reader is referred their sections 3.3 and 3.5 for a detailed derivation.

\begin{figure}
  \centering
  	\includegraphics[width=1 \textwidth]{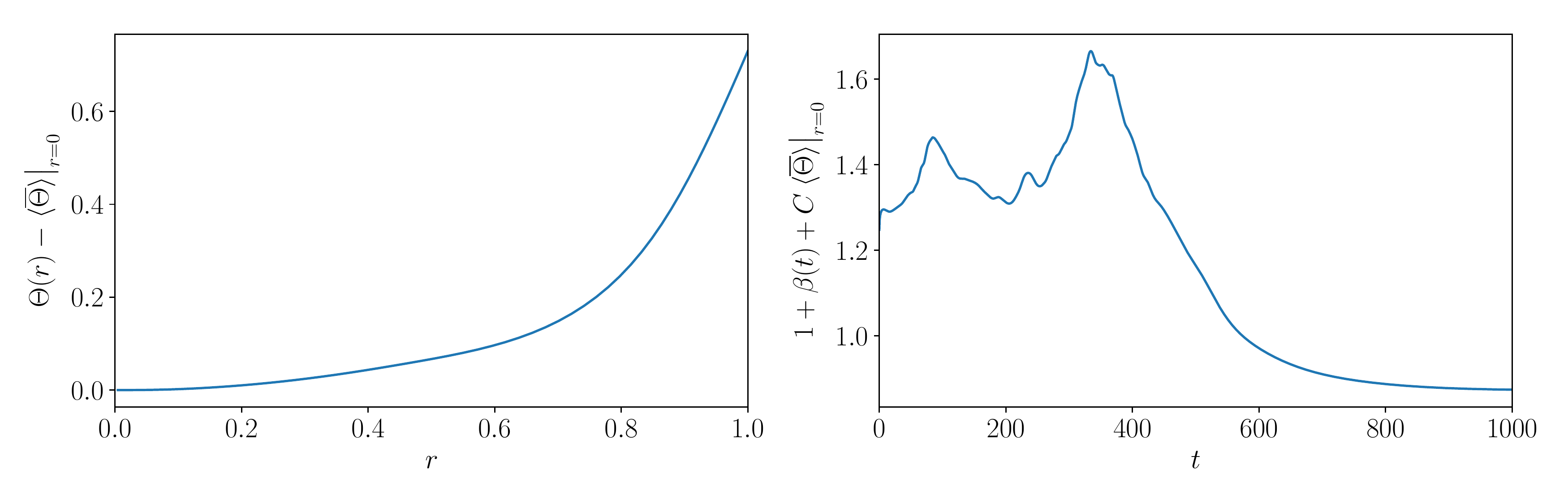}
  \caption{\small Application of HHS's relaminarisation criterion \eqref{eq:hss-3.12} in the case $C=2$ and $Re=3000$. Left: Temperature profile shifted by $\left.\langle\overline{\Theta}\rangle\right|_{r=0}$. Right: the corresponding pressure gradient.}
  \label{fig-HHS-1}
\end{figure}
\begin{figure}
  \centering
  	\includegraphics[width=1 \textwidth]{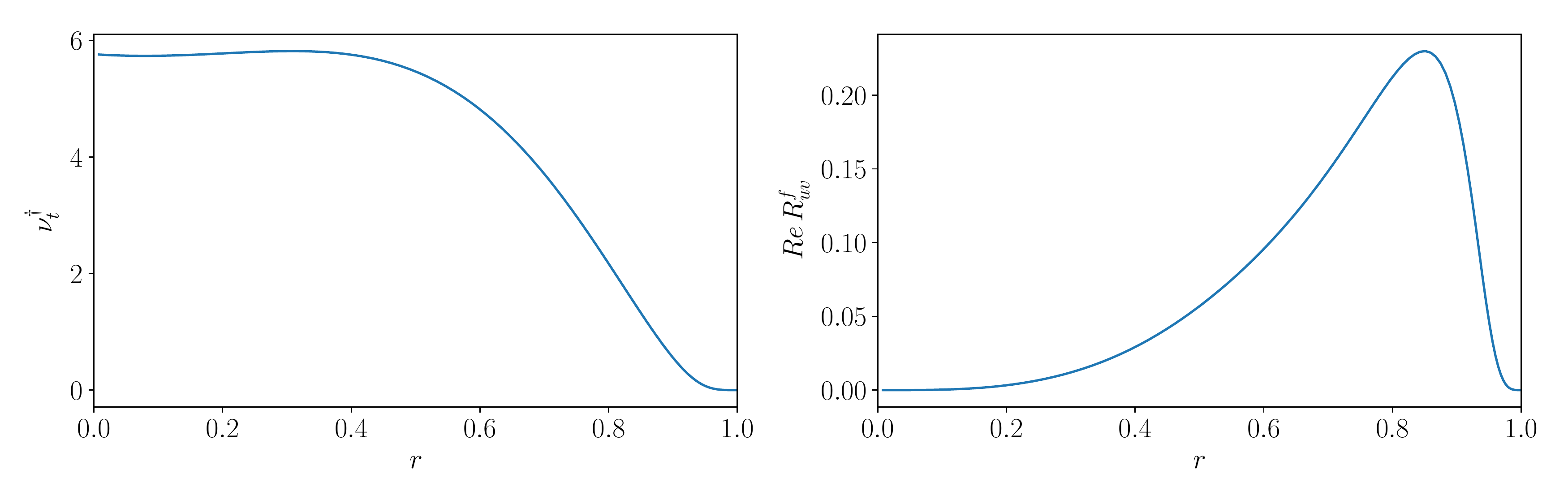}
  \caption{\small Eddy viscosity (left) of the EPG flow and Reynolds shear stress (right) of the body-force perturbed flow in the case $C=2$ and $Re=3000$. The eddy viscosity is calculated following an approach similar to \cite{WHC10}, as summarised in Appendix \ref{sect:Cess}. \Rtwo{Once $\nu_t^{\dagger}$ is known, $\mathscr{R}_{uv}^f(r)$ is calculated using \eqref{eq-3.7-HHS}, together with \eqref{eq-3.6-HHS}.}}
  \label{fig-HHS-2}
\end{figure}

Here, we apply the criterion for relaminarisation \eqref{eq:hss-3.14} proposed by HHS to our model for a vertical heated pipe. The radially dependent 
body-force is $f_0=(4C/Re) \langle\overline{\Theta}\rangle(r)$.
 Since the body-force in HHS is zero at the axis, we shift the temperature profile by its value at the axis $\left.\langle\overline{\Theta}\rangle\right|_{r=0}$ and absorb this constant into the pressure gradient (see figure \ref{fig-HHS-1}).  This leads to the body force
\begin{equation}
  f_1(r)=(4C/Re) \Big[\langle\overline{\Theta}\rangle-\left.\langle\overline{\Theta}\rangle\right|_{r=0}\Big]
\label{eq-f1}
\end{equation}
 and a fixed-pressure Reynolds number 
 \[
 Re_p=Re\Big[(1+\beta) + C\left.\langle\overline{\Theta}\rangle\right|_{r=0}\Big]
 \]
Initially, we consider the simulation with $C=2$ and $Re=3000$ for which it is observed 
that $Re_p=4252.71$.
By inserting $f=f_1$ in $\mathcal{I}_1$ we obtain $Re\,\mathcal{I}_1=0.12$.
To calculate $\mathcal{I}_2$ we need to evaluate 
the EPG flow in order to obtain $\nu_{t}^{\dagger}(r)$ \Rtwo{and hence the Reynolds stress term $\mathscr{R}_{uv}^f(r)$ via \eqref{eq-3.7-HHS} and \eqref{eq-3.6-HHS}}.
 By definition, $Re_p^{\dagger}=Re_p$.
In an approach similar to \cite{WHC10}, summarised in Appendix \ref{sect:Cess}, 
the eddy viscosity $\nu_{t}^{\dagger}(r)$ of the EPG reference flow is calculated
 using an expression originally suggested by \cite{Cess1958}, \Rtwo{see \eqref{eq:CessEy}}. 
 The resulting eddy viscosity is shown in figure \ref{fig-HHS-2}(left). 
\Rtwo{By substituting $\nu_{t}^{\dagger}$ in \eqref{eq-3.6-HHS} we can invert for $\mathrm{d} \mathscr{U}_z^f/\mathrm{d}r$ which plugged into \eqref{eq-3.7-HHS}
gives us the Reynolds stress $\mathscr{R}_{uv}^f(r)$ (see figure \ref{fig-HHS-2}(right)). Finally, by inserting the latter in the second integral of \eqref{eq:hss-3.12} we obtain $Re\,\mathcal{I}_2=0.0405$.}
Putting everything together, (\ref{eq:hss-3.12}) gives
\Rtwo{ $U_b^f=Re\mathcal{I}_1+Re\mathcal{I}_2=0.12+0.0405\approx0.16$}. Then, \Rtwo{using \eqref{eq:hss-3.14}}, $Re_{app} = Re\,\left(1-2\,U_b^f\right)=2040 < 2300$, i.e. the flow 
is expected to relaminarise.
This value obtained for the apparent Reynolds number is reasonable, since relaminarisation occurs after approximately 400 time units (see figure \ref{fig-HHS-1} (right)).

\subsection{HHS prediction of phase diagram and nonlinear dynamics}
\label{sec:HHSpred}

We now consider the general case of a flow at $Re$ with heating $C$, while introducing a number of approximations to simplify the analysis. 

Firstly, the case discussed in \S \ref{sec:HHS} \Rtwo{($C=2$ and $Re=3000$)} suggests that $Re\,\mathcal{I}_1$ has a significantly greater contribution than $Re\,\mathcal{I}_2$ in determining the body-force perturbation flow. This is found to be generally true for the cases considered herein, as well as those discussed in HHS, and hence we omit the term $Re\,\mathcal{I}_2$ for simplicity below.
The perturbation flow due to the body force can thus be evaluated as
\begin{equation}
U_b^f \approx Re\, \mathcal{I}_1 = \frac{1}{2} Re \int_0^1 (1-r^2)f(r)\,r\mathrm{d}r = 2C \int_0^1(1-r^2) 
\Big[\langle\overline{\Theta}\rangle-\left.\langle\overline{\Theta}\rangle\right|_{r=0}\Big]\mathrm{d}r\,,
\label{eq-Ubfapprox}
\end{equation}
where \eqref{eq-f1} has been used for $f(r)$.

Secondly, figure \ref{fig:sim5300}(bottom right) shows that the temperature mean profiles are remarkably similar in all turbulent shear-driven flows (i.e.\ ignoring the 
laminar or convection driven flow states), as far as 
the integral part of the right-hand side of (\ref{eq-Ubfapprox}) is concerned, 
despite that the values of the $\Nu$ (proportional to the gradient at the wall) are necessarily quite different for different cases. 
For the case $Re=5300$ $C=3.75$, for the left-hand side of \eqref{eq-Ubfapprox} we obtain $Re\,\mathcal{I}_1 = 0.164$. By applying the above assumption, 
\begin{equation}
U_b^f \approx Re\,\mathcal{I}_1=\frac{0.164}{3.75}\,C=0.04\,C
\end{equation}
Let $Re_{app}$=2300 to find the critical $C$ for flow laminarisation, that is,
\begin{equation}
Re(1-2U_b^f) = Re(1-0.08\,C)=2300
\end{equation}
or
\begin{equation}
C_{cr,1}=12.5\left(1-\frac{2300}{Re}\right) \, .
\label{eq-Clam}	
\end{equation}
For $C\gtrsim C_{cr,1}$ we expect to see rapid transition from the 
shear-driven turbulent state to the convective state.
Noting $C=Gr/(16Re)$, the above can be expressed as a critical Grashof number:
\begin{equation}
Gr_{cr,1}=200(Re-2300)
\label{eq-Grlam}	
\end{equation}

Let us now consider the opposite scenario in which the flow
under heating $C$ is either laminar or convection driven. Figure \ref{fig:sim5300}(bottom right) shows that the temperature profiles in such flows are significantly different from those in a turbulent shear-driven flow, and generally with a much thicker thermal boundary layer, and hence a greater buoyancy force. Consider the extreme case when 
the radial heat transfer is purely due to conduction and the temperature distribution is given by $\langle \overline{\T}\rangle=r^2$. The buoyancy-driven perturbation flow is therefore
\begin{equation}
U_b^f \approx Re\, \mathcal{I}_1 = 2C \int_0^1(1-r^2)\, r^2\, \mathrm{d}r =\frac{C}{6}
\end{equation}
Then a second critical $C=C_{cr,2}$ can be evaluated, 
\begin{equation}
C_{cr,2}=6\left(1-\frac{2300}{Re}\right) \, ,
\label{eq-Cturb}	
\end{equation}
below which the flow is expected to transition to the shear-driven turbulent flow. 
To put it another way, 
it is predicted that metastability of the shear-driven turbulent state 
should {\em not} be observed for $C\lesssim C_{cr,2}$, so that
the turbulent state is stable.
Between $C_{cr,1}$ and $C_{cr,2}$ the shear-driven state is expected to be
metastable, so that this or a convective state may be observed.
In terms of the Grashof number,
\begin{equation}
Gr_{cr,2}=96(Re-2300).
\label{eq-Grturb}		
\end{equation}

\begin{figure}
   \centering
	\includegraphics[width=0.9\textwidth]{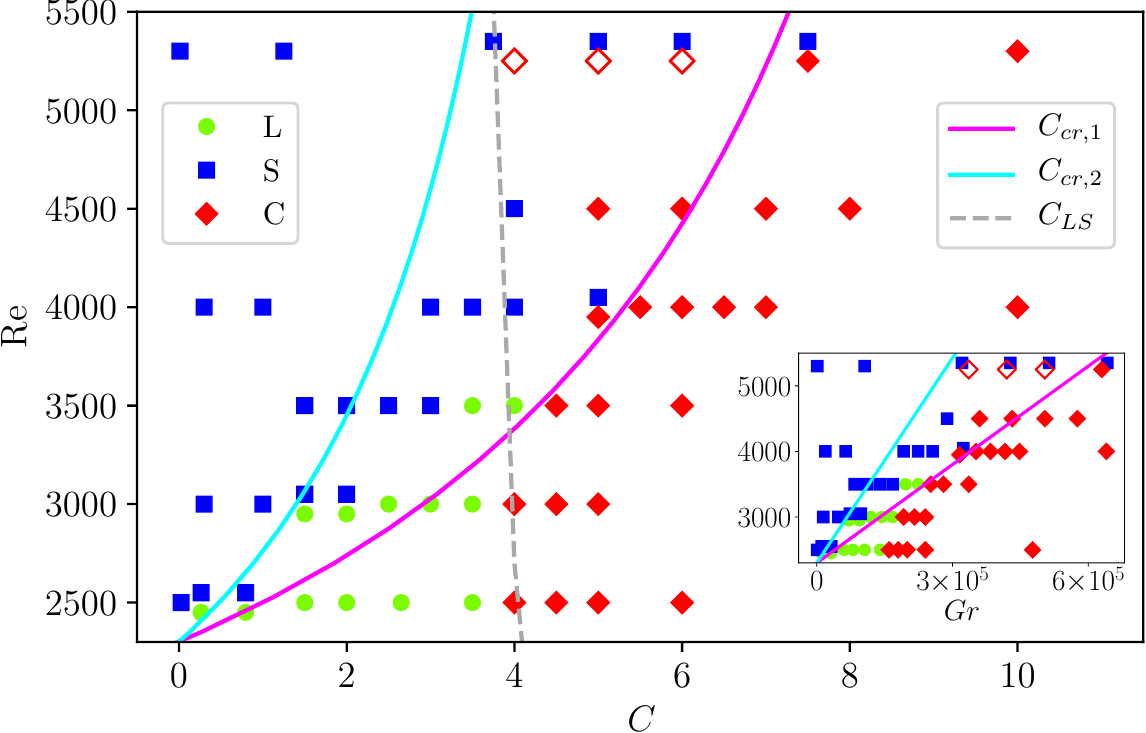}
  \caption{\small  Regions of laminar (L) flow, shear-driven (S) turbulence and 
   convection-driven (C) flow, as in \ref{fig:LTconv}, together with equations \eqref{eq-Clam} and \eqref{eq-Cturb} and the linear stability stability curve
   \Rtwo{(dashed red curve in figure \ref{fig-heatedpipe-LS-1})}.  
   Initial conditions are a shear-driven turbulent state, except 
   for the hollow symbols at $Re=5300$ which are started with a convection driven state,
   and similarly cases towards the bottom-right, where it is clear that the shear-driven
   state decays immediately.
}
  \label{fig-LTvsGr}
\end{figure}

Equations \eqref{eq-Clam} and \eqref{eq-Cturb} are plotted on the $Re-C$ graph in figure \ref{fig-LTvsGr} together with all DNS results already presented in figure \ref{fig:LTconv}. 
The data of figure \ref{fig:LTconv} was obtained starting from shear-driven turbulent states. 
Some additional simulations were performed at $Re=5300$ starting from convection-driven states and are reported in figure \ref{fig-LTvsGr} using hollow symbols, with a slight offset in $Re$ for visualisation reasons. Note that in a $Re-Gr$ graph \eqref{eq-Grlam} and \eqref{eq-Grturb} are straight lines  (see the inset in figure \ref{fig-LTvsGr}).


 Considering a series of DNS runs for a fixed $Re$, for example $Re=5300$, but increasing $C$ values (heating) starting from $C=0$, equation \eqref{eq-Clam} gives the critical $C=C_{cr,1}$ above which the flow will be laminarised or switch to convection-driven. On the other hand, starting from a large $C$ when the flow is laminarised or convective, equation \eqref{eq-Cturb} predicts a critical $C=C_{cr,2}$ 
below which the flow will be turbulent when sufficient disturbances are provided in the DNS. 
As $C_{cr,1}$ is larger than $C_{cr,2}$ for a given $Re$,
there is an overlap in the possible state of flow,
and consequently there is a hysteresis region
in which the flow may or may not be laminarised, depending on the initial flow of the simulation (or experiment). As a result, the $Re-C$ plane can be divided into three regimes by the curves representing the two equations, i.e., turbulent shear-driven flow (regime I), convection-driven or laminar flow (regime III) and regime II in which either of the above may happen dependent on the initial flow. Note that for the Reynolds number range considered here, the linear stability curve (showed as a dashed grey line in figure  \ref{fig:LTconv}) is always to the right of $C_{cr,2}$, i.e. $C_{cr,2}<C_{LS}$ . The two curves cross at $Re\approx6000$ (not shown), which means that, for $Re<6000$ the convective flow is always linearly stable if $C < C_{cr,2}$. Hence, below $Re\approx6000$, shear driven turbulence may be observed for $C<C_{LS}$.

\begin{figure}
\centering
\includegraphics[width=0.9\textwidth]{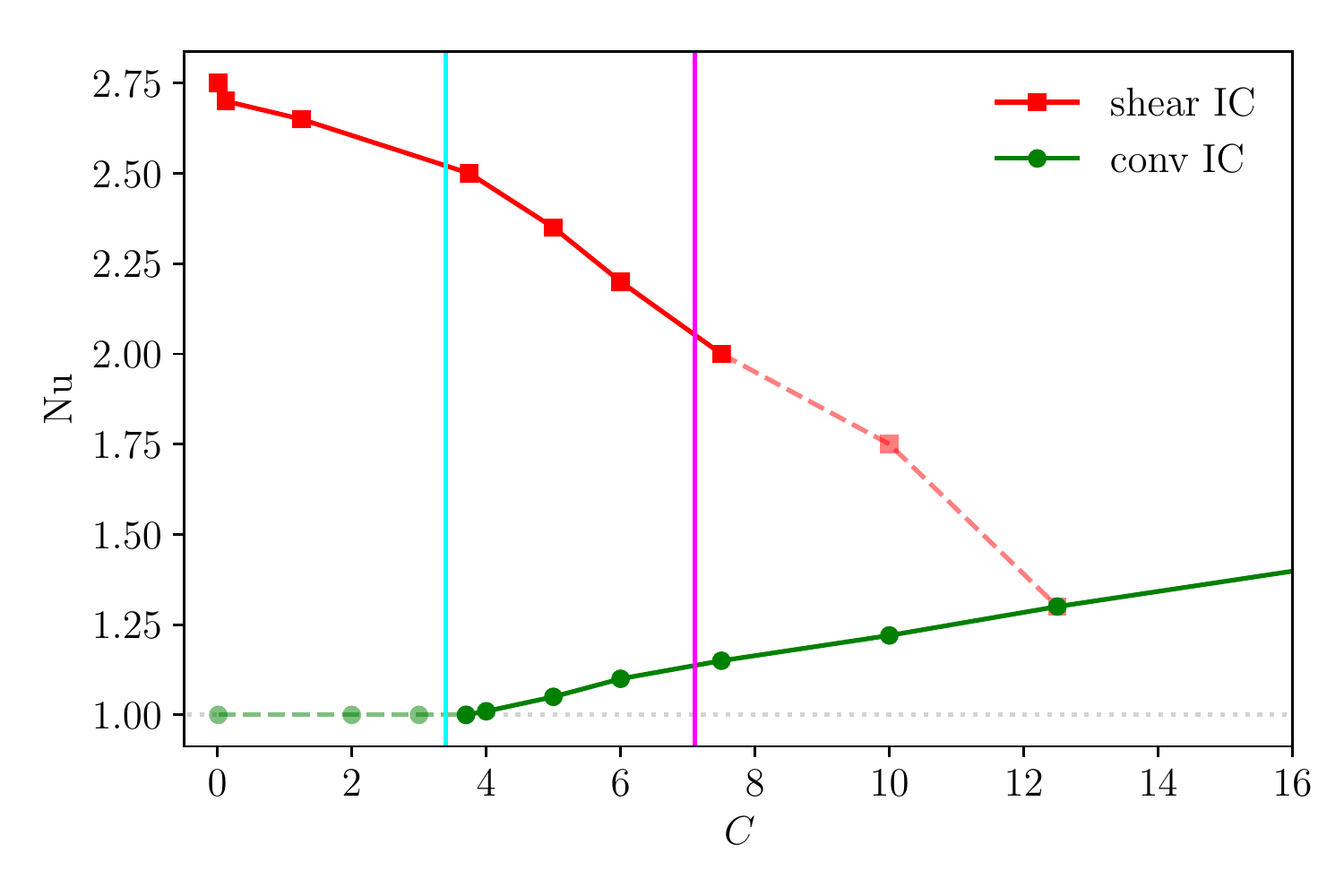}
\caption{ \label{fig:NuvsC}
Nusselt number vs $C$ for simulations started with shear and convection initial conditions at $Re=5300$.
\Rtwo{ The magenta and cyan vertical lines correspond to the critical buoyancy parameters $C_{cr,1}$ and $C_{cr,2}$ given by \eqref{eq-Clam} and \eqref{eq-Cturb}, respectively. For values of $C \gtrapprox C_{cr_1}$ ($C \lessapprox C_{cr_2}$) the shear-driven (convection-driven) state is not supported and correspondingly the upper (lower) branch is plotted with a dashed semi-transparent line}.
}
\end{figure}

\begin{figure}
\centering
\includegraphics[height=0.9\textwidth]{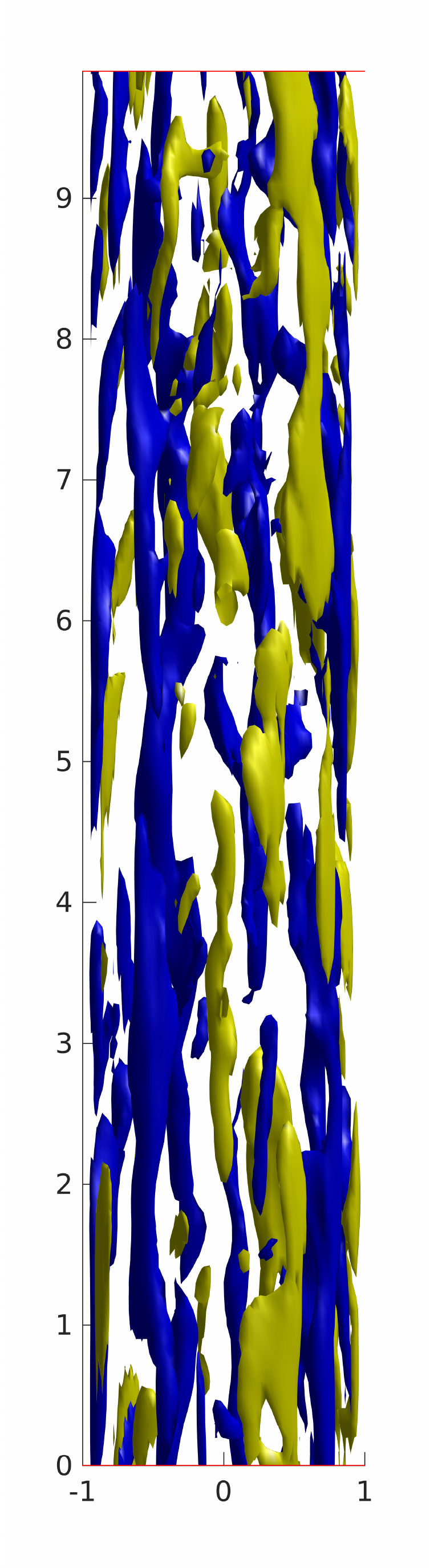} \quad 
\includegraphics[height=0.9\textwidth]{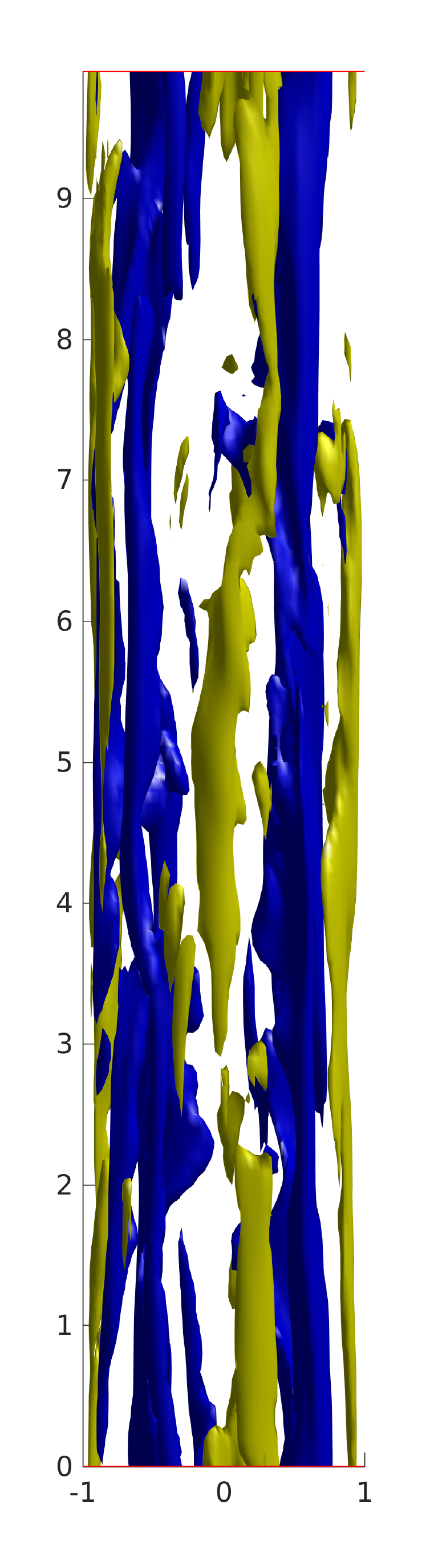} \quad
\includegraphics[height=0.9\textwidth]{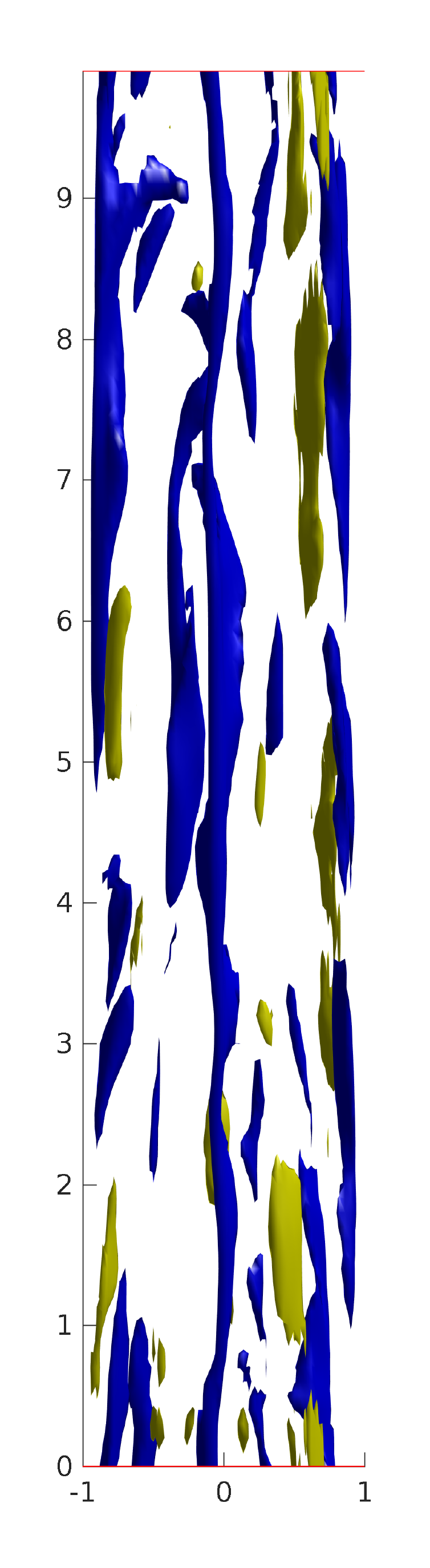} 
\caption{ \label{fig:3dvis-streaks} Three-dimensional visualisations of low (blue) and high (yellow) speed streaks in the isothermal (left), heated (middle) and EPG (right) flows. Isosurfaces of turbulent streamwise velocity normalised by the corresponding apparent friction velocity $u_z'/u_{\tau p}=\pm 4$. }
\end{figure}

\begin{figure}
\centering
\includegraphics[height=0.9\textwidth]{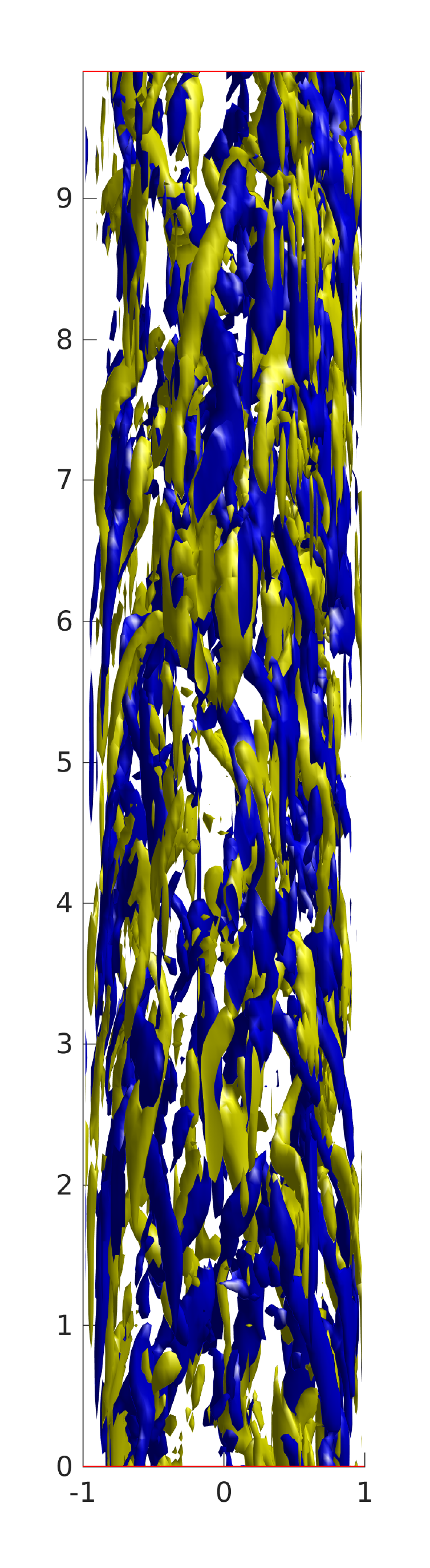} \quad 
\includegraphics[height=0.9\textwidth]{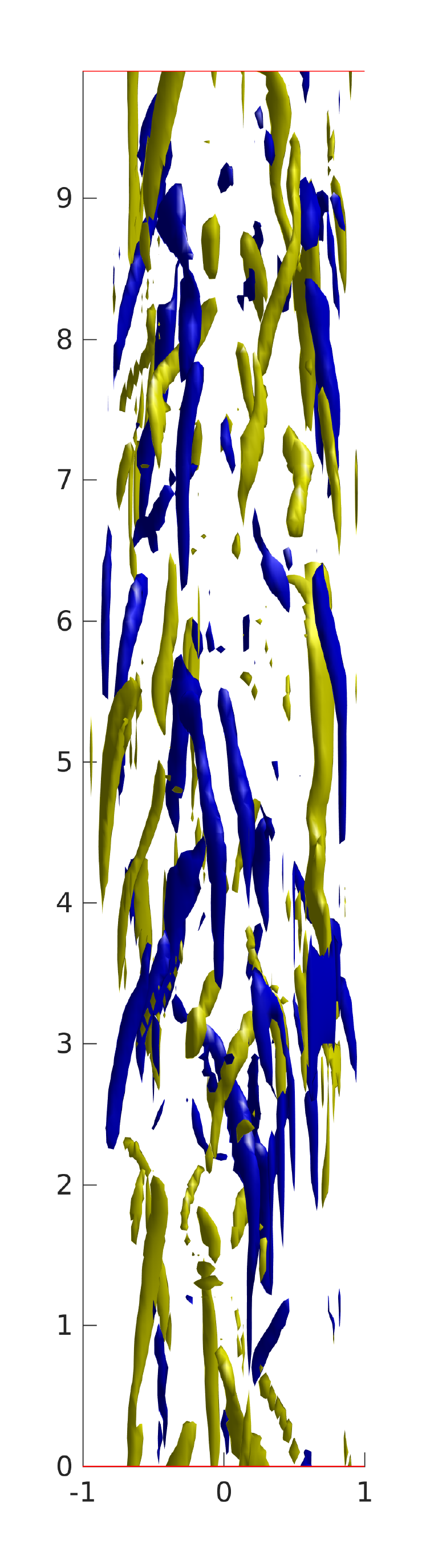} \quad
\includegraphics[height=0.9\textwidth]{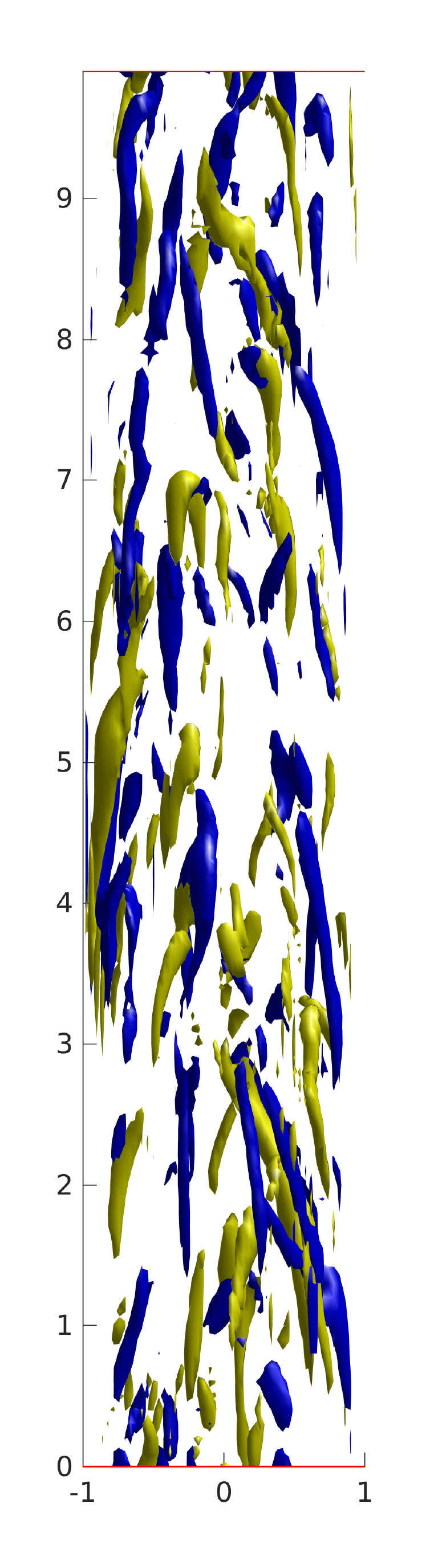}
\caption{ \label{fig:3dvis-vort} Three dimensional visualisations of vortical structures in the isothermal (left), heated (middle) and EPG (right) flows. Isosurfaces of streamwise vorticity fluctuations normalised by the corresponding apparent friction velocity $\omega_z'/u_{\tau p}=\pm 35$. }
\end{figure}

A plot showing the phase transitions
for the fixed Reynolds number $Re=5300$ is provided in figure \ref{fig:NuvsC}, 
where the Nusselt number is displayed as a function of $C$ for simulations started with either shear-driven or convection-driven states.
 The two critical $C$ at this Reynolds number, $C_{cr,1}=7.1$ and $C_{cr,2}=3.4$, are indicated with vertical lines in figure \ref{fig:NuvsC}.
 Starting from an unheated ($C=0$) turbulent flow, applying a low heating ($C \lessapprox 7$), we observe that the flow remains turbulent over the entire period of simulation \Rtwo{($t = 2000$)}. The dynamics thus sits on the upper branch shown in figure \ref{fig:NuvsC}. As $C$ is increased, the lifetime of shear-turbulence 
 drops 
 below 2000 time units for $C \gtrapprox 7.5$ and turbulence only survives for less than 500 time units at $C = 10$. It then switches to the convection-type flow. This behaviour is marked in figure \ref{fig:NuvsC} by plotting the upper-branch curve with a dashed line for $C \ge 7.5$ 
 until it crosses the lower-branch at $C = 12.5$. 
  At this value of $C$, indeed, the switch to the convective flow appears to be immediate. Now, starting from this convection-driven flow and applying a lower $C$, the flow remains convection-driven turbulent for $C \ge 3.8$, or relaminarises for $C \lessapprox 3.8$.  
This value of $C$ corresponds to the onset of the linear instability, which is 
responsible for the kink in $\Nu$ as $C$ is decreased.
Our previous analysis predicts that for flows on the left of \eqref{eq-Cturb}, their $Re_{app}$ 
is greater than 2300, hence they may be prone to transition to turbulence subject to sufficient disturbances.
  Correspondingly, the lower-branch curve in figure \ref{fig:NuvsC} is plotted with a dashed line for $C< C_{cr,2} = 3.4$ to indicate that in practice (e.g. in a lab experiment) the flow would become shear-driven turbulent again. 
However, as previously discussed, at this Reynolds number, $C_{cr,2}<C_{LS}$.
  Bistability (between shear or convection driven states) is thus observed for $3.8 \lessapprox C \lessapprox 7.5$.
The latter value is in very good agreement with the threshold $C_{cr,1}=7.1$ predicted above. 
In figures \ref{fig:3dvis-streaks} and \ref{fig:3dvis-vort} the turbulent structures of the isothermal and heated flows at $Re=5300$, $C=0$ and $5$, are compared to those of the EPG reference flow. The latter was computed by performing a DNS with fixed pressure gradient such that $Re_p^{\dagger}=Re_p=10898.7$.
 The flow structures - streaks and vortices - are visualised as isosurfaces of streamwise velocity and streamwise vorticity fluctuations, normalised by the apparent friction velocity based on the pressure gradient component of the wall shear stress only,
$u_{\tau p}^*$, where the asterisk $^*$ denotes a dimensional quantity here.
The resulting apparent friction Reynolds number is $Re_{\tau p}:=u_{\tau p}^* R^*/\nu^* = Re_{\tau}^{\dagger}=147.6$.

\begin{figure}
\centering
(a)\hspace{50mm}(b) \\
\includegraphics[width=0.48\textwidth]{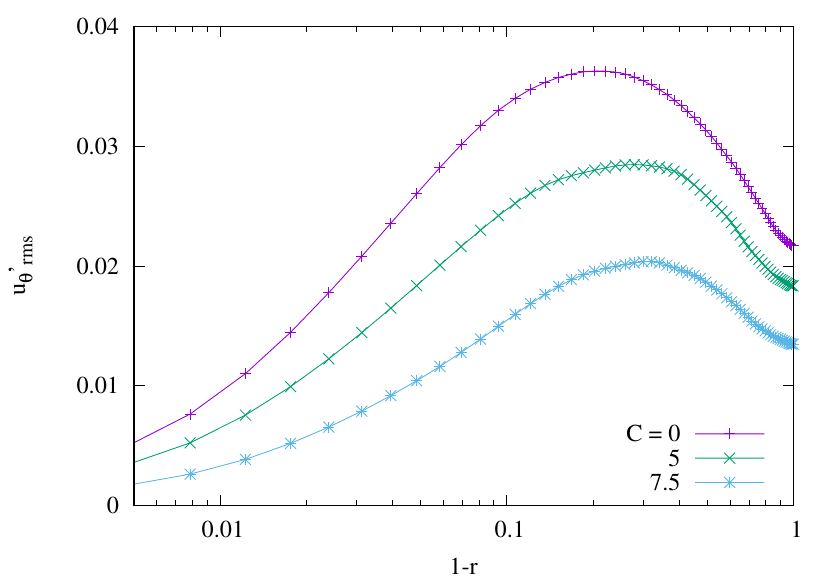}  
\includegraphics[width=0.48\textwidth]{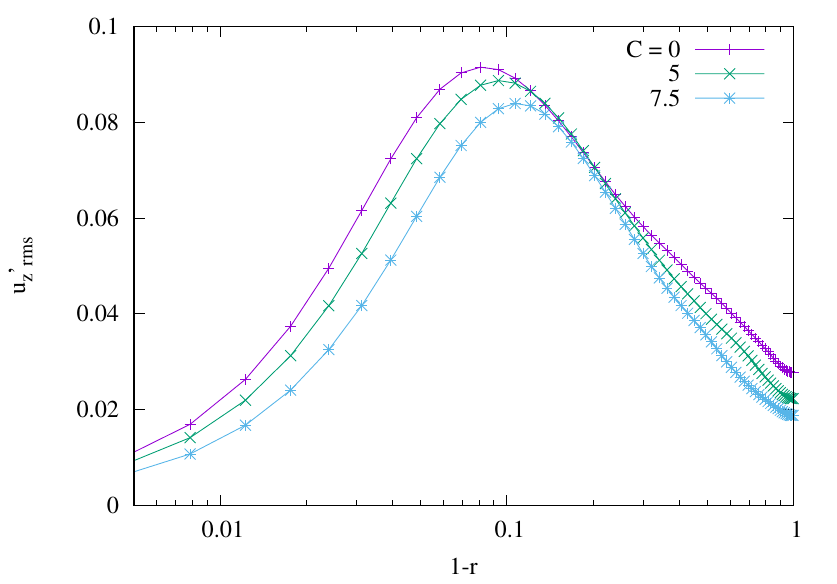}\\
(c)\includegraphics[width=0.48\textwidth]{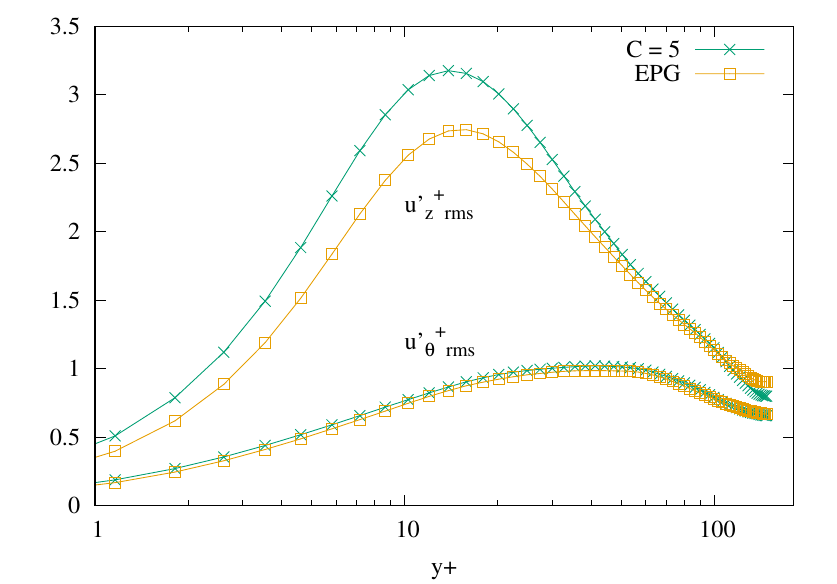}\\
\caption{ \label{fig:RMSu'} \Rtwo{RMS velocity fluctuations.
(a) $u'_\theta$, a measure of `rolls', are suppressed as $C$ increases,
while (b) $u'_z$ a measure of `streaks', are little changed.  
(c) rolls for the $C=5$ case correspond closely to its EPG counterpart,
while the heated case has slightly stronger streaks.}}
\end{figure}

Comparison between the isothermal and heated flows show that the streaks are 
relatively unaffected, while vortices are significantly weakened. 
Our interpretation is that while the streaks are responsible
for the saturation of the nonlinearity of the flow, via nonlinear
normality of the mean flow \citep{W95b}, it is relatively `easy'
to produce streaks.
Note that the mean axial flow for these cases is almost identical 
(figure \ref{fig:sim5300}), and at the end of \S \ref{sec:N4L} large initial
amplifications of disturbances remains possible in the heated case.
It is observed that weaker vortices in the heated case are sufficient
to produce saturated streaks of the same amplitude.  
Thus, vortices are more important in the sense
that criticality for transition appears to occur 
when the vortices are too weak. 
Comparing now the heated flow with the EPG flow, 
consistent with the observations of HHS (see their figure 19), it can be seen that the streaks 
\Rtwo{in the heated flow are typically stronger than in the EPG flow, 
while the vortices are of similar strength.
In figure \ref{fig:RMSu'} we plot RMS velocity fluctuations.  
Axial perturbations (a) are not strongly affected by the heating, while the
cross-flow components (b) are significantly suppressed. 
(The plot for $u'_r$ is very similar to that shown for $u'_\theta$.)  
In (c) it is seen that the heated and EPG flow have very similar 
cross-components, while axial perturbations in the heated case are slightly 
stronger than in the EPG flow.  
These results are consistent with observations from the 
three-dimensional visualisations of figures \ref{fig:3dvis-streaks} and
\ref{fig:3dvis-vort}, and likewise suggest that it is the weakening
of rolls rather than streaks that appear to be responsible for 
laminarisation.
}


\section{Conclusions}
\label{sec:conclusions}
In this paper
 we have studied the flow of fluid through a vertically-aligned heated pipe using direct numerical simulations (DNS),
  linear stability 
 and nonlinear travelling-wave solution analyses.
The flow is driven by an externally applied pressure gradient and aided by the buoyancy resulting from the lightening of the fluid close to the heated wall.
DNS were performed for a range of Reynolds numbers $Re$ and buoyancy parameters $C$, where the latter measures
 the magnitude of the buoyancy force relative to the the pressure gradient of the laminar isothermal shear flow, 
and three different flow regimes were identified -- laminar flow, shear-driven turbulence and convection-driven flow -- depending on the flow parameters.
At relatively low $Re \lesssim 3500$ turbulence is completely suppressed (relaminarised) by buoyancy and as $C$ is increased convection starts driving a relatively quiescent flow. For larger $Re$, instead, the shear-driven turbulent flow transitions directly to the convection-driven state.
Consistent with the appearance of the convective state observed in simulations,
a linear instability was found at $C \approx 4$, roughly independent of $Re$ for most of the
range considered.
The result of increasing $C$ can be compared to that of 
increasing polymer concentration, or Weissenberg number $Wi$,
which is known to have a drag reducing effect on turbulent flows \citep{virk-etal-1967}.
Similar to our phase diagram (figure \ref{fig:LTconv}), 
\Rthree{a region of relatively quiescent flow has been reported
for a certain range 
 of $Re$ and $Wi$}
\citep{choueiri-etal-2018,lopez-etal-2019},
\Rthree{although the underlying physical mechanism 
(elastoinertial instability) is clearly very different from the one studied here (convection driven).}

Cases where turbulence is suppressed exhibit a flattened mean streamwise velocity profile.
In agreement with recent observations by \citet{kuhnen-etal-2018a} and \citet{marensi-etal-2019} on the effect of flattening,
we found that states that mediate turbulence (lower-branch travelling wave solutions) are ``pushed out'' from the laminar state, i.e. as $C$ increases, a larger perturbation amplitude or larger $Re$ are required to drive shear turbulence until, for sufficiently large $C$, the travelling wave is suppressed altogether.
Finally, we used the relaminarisation criterion recently proposed by \cite{he-etal-2016}, based on an ``apparent Reynolds number'' of the flow,
to predict the critical $C=C_{cr,1}(Re)$ above which the flow will be laminarised or switch to the convection-driven type. 
This apparent Reynolds number is 
based on an apparent friction velocity
 associated with only the pressure force of the flow (i.e. excluding the contribution of the body force/buoyancy).
Bistability between shear or convection-driven states was found to occur in the region $4 \lesssim C \lesssim C_{cr,1}$ where
the flow may or may not be laminarised depending on the initial flow of the simulation or experiment.

Comparison of the turbulent flow structures (rolls and streaks) with those of two reference flows - the flow of equivalent pressure gradient (EPG) and that of equivalent mass flux (EFR) - suggests that near criticality for relaminarisation the vortices, rather than the streaks, are more important in the sense that criticality for transition occurs when the vortices are too weak.
\Rtwo{This picture is not straight forward to reconcile with the interpretation
of \cite{kuhnen-etal-2018a}, where relaminarisation is attributed to reduced ability to produce streaks in the presence of the flattened base
profile.  
In the heated case, 
the base velocity profile does not appear to change significantly 
while shear-driven turbulece is present.
Thus it appears unlikely that transient growth of streaks 
is affected by the heating.
Indeed, laminarisation occurs despite little suppression of the streaks.
The experiments of \cite{kuhnen-etal-2018a} are slightly different, however,
in that the various flow manipulations they introduce 
{\em do} 
change the base profile of the flow.  
In that case it is correct that transient growth will be affected, 
although we conjecture that
it is the suppression of the vortices due to suppression of the streaks 
that is responsible for laminarisation in that case.  
Their numerical
experiments in the presence of a force are very similar to the calculations 
here and of HHS.  
In that case we expect the mechanism we have described to be more 
clearly responsible for the laminarisation.
}\\\\






{\bf{Acknowledgments.}} \Rtwo{The anonymous referees are kindly acknowledged for their useful suggestions and comments.}\\

{\bf{Funding.}} This work was funded by EPSRC grant EP/P000959/1.\\

{\bf{Declaration of Interests.}} The authors report no conflict of interest.\\

\appendix

\section{Link between upward-heated and downward-cooled cases}
\label{sect:link}
 Consider the axial force from
the pressure gradient and buoyancy terms in (\ref{eq:veleq}).
Ignoring the factor $4/Re$ that multiplies all terms, let
\begin{equation}
   \label{eq-A1}
   1 + \beta + C\,\T = 1 + \tilde{\beta} + \tilde{C}\,\tilde{\T}\, .
\end{equation}
with $C>0$ for the upward heated case on the left hand side.
Let the right hand side represent the downward cooled case, taking
$\tilde{\T}=1-\T$ so that $\tilde{\T}$ is coolest on the boundary 
($\tilde{\T}=1-r^2$ for the laminar case).  Put $\tilde{C}=-C<0$, 
as buoyancy due to positive temperature variations oppose the pressure gradient.
(Cooling, however, aids the downward flow.)
Substituting in \eqref{eq-A1} we find 
$\tilde{\beta}=\beta+C$, i.e.\ the systems differ only 
by a known offset in the pressure gradient required to maintain volume 
flux.

\section{Turbulent base flow and eddy viscosity}
\label{sect:Cess}

%
The turbulent mean flow profile for a pipe may be written
$\vec{U}=U(y)\hat{\vec{z}}$, where
$y=1-r$ is the dimensionless distance from the boundary wall and $r$ is the radial coordinate.
%
%
Applying the Boussinesq eddy viscosity to model for the
turbulent Reynolds-stresses,
the streamwise component of the Reynolds-averaged momentum
conservation reads
\begin{equation}
   \label{eq:avgedNS}
   \frac{1}{Re}
   \left(
      \frac{1}{r} + \pd{r}
   \right)
   (\nu_T \pd{r} U )  = \pd{z}P ,
\end{equation}
where the total effective viscosity is $\nu_T(y)=1+\nu_{t}(y)$
and $\nu_{t}$ is the eddy-viscosity, normalised such that $\nu_T(0)=1$,
i.e.\ the kinematic value is attained at the wall.

To calculate $\nu_{t}$ it is convenient to use the expression originally suggested for pipe flow
by \cite{Cess1958}, later used for channel flows by  
\cite{Reynolds1967} and then by many others
\citep{Butler1993,delAlamo2006,Pujals2009}:
\begin{equation}
   \label{eq:CessEy}
   \nu_{t}(y) =  \frac{1}{2} \left\{
            1 + \frac{\kappa^2 \hat{R}^2 \hat{B}}{9}
      \left(2y-y^2\right)^2
      \left(3-4y+2y^2\right)^2
      \left[
         1 - e^{\frac{-y \hat{R} \sqrt{\hat{B}}}{A^+}}
      \right]^2\right\}
  ^{\frac{1}{2}} - \frac{1}{2} \, .
\end{equation}
Here, $\hat{R}=Re\,/\,2$,\, $\hat{B}=2\,B$, with $B = -\pd{z}P$ being
the averaged streamwise pressure gradient.
The parameters $A^+=27$ and $\kappa=0.42$ 
have been chosen to fit the more recent observations of \citep{McKeon2005}.

For the calculation of \S\ref{sec:HHS}, the (apparent) pressure gradient $B$ 
and 
(apparent) $Re_p$ are known. 
The mass flux $Re$ of (\ref{eq:avgedNS}) 
is not yet known, and we wish to determine $\nu_t$.
An initial estimate for $Re$ is obtained from the approximation of \cite{blasius-1913}, which may be written
\begin{equation}
   Re_p = \frac{0.0791}{16}\,Re^{1.75}\,.
\end{equation}
Then, (\ref{eq:CessEy}) can be used to calculate
$\nu_t(r)$, but we must check consistency with (\ref{eq:avgedNS}).  The latter
equation can be inverted for $U(r)$, and, as it has been non-dimensionalised with the 
same scales of section \S\ref{sect:nondim}, the mean velocity
$U_b=2\int_0^1 U(r)\,r\,\mathrm{d}r$ should be $0.5$.  It will not be exactly so, 
as 
$Re$ (for the given $\partial_zP$)
has only been estimated.  A better estimate 
is given by $Re := (0.5/U_b)\,Re$, so that $\nu_t$ can be recalculated and 
iteratively improved.


\clearpage


\bibliographystyle{jfm}
\bibliography{./pipes}

\end{document}